\documentclass[11pt]{article}
\usepackage[utf8]{inputenc}
\usepackage{setspace}

\usepackage[margin=1in]{geometry}

\usepackage{apacite}
\usepackage{blindtext}
\usepackage{amsmath,bm}
\usepackage{color}
\usepackage{rotating}

\usepackage{caption}
\usepackage{subcaption}

\usepackage{grffile}

\usepackage{graphicx}

\usepackage{lscape}

\usepackage{blindtext}

\usepackage{amssymb}
\usepackage{varioref} 
\usepackage{textcomp}

\usepackage{booktabs}

\usepackage{multirow}

\usepackage{pdflscape}

\usepackage{graphicx}

\usepackage{relsize}
\usepackage{tablefootnote}
\usepackage{wrapfig}
\usepackage{natbib}
\usepackage{listings}
\usepackage{longtable}
\usepackage{url}
\usepackage{tabularx}
\usepackage{float}
\usepackage{epsfig}
\usepackage[titletoc]{appendix}
\usepackage{arydshln}
\usepackage{footnote}
\usepackage{secdot}
\usepackage{bigints}
\usepackage[utf8]{inputenc}
\usepackage[english]{babel}
\usepackage{xr}
\usepackage{adjustbox}
\usepackage{graphicx}

\usepackage{multicol}
\usepackage{enumitem}

\usepackage{bigstrut}
\usepackage{rotating}
\usepackage{xr}

\title{A Stochastic Version of the EM Algorithm for Mixture Cure Rate Model with Exponentiated Weibull Family of Lifetimes}
\author{S. Barui\thanks{S. Barui is with Quantitative Methods and Operations Management Area, Indian Institute of Management, Kozhikode, Kerala, India.} , S. Pal\thanks{S. Pal is with Department of Mathematics, University of Texas at Arlington,  Texas, USA (email: suvra.pal@uta.edu).}, N. Mishra\thanks{N. Mishra is with Department of Mathematics and Statistics, University of South Alabama,  Alabama, USA}  and K. Davies \thanks{K. Davies is with Department of Statistics, University of Manitoba, Winnipeg, Canada.} }
\date{}

\begin{document}
\maketitle
\begin{abstract}
Handling missing values plays an important role in the analysis of survival data, especially, the ones marked by cure fraction. In this paper, we discuss the properties and implementation of stochastic approximations to the expectation-maximization (EM) algorithm to obtain maximum likelihood (ML) type estimates in situations where missing data arise naturally due to right censoring and a proportion of individuals are {\it immune} to the event of interest.  A flexible family of three parameter exponentiated-Weibull (EW) distributions is assumed to characterize lifetimes of the {\it non-immune} individuals as it accommodates both monotone (increasing and decreasing) and non-monotone (unimodal and bathtub) hazard functions. To evaluate the performance of the SEM algorithm, an extensive simulation study is carried out under various parameter settings. Using likelihood ratio test we also carry out model discrimination within the EW family of distributions. Furthermore, we study the robustness of the SEM algorithm with respect to outliers and algorithm starting values. Few scenarios where  stochastic EM (SEM) algorithm outperforms the well-studied EM algorithm are also examined in the given context. For further demonstration, a real survival data on cutaneous melanoma is analyzed using the proposed cure rate model with EW lifetime distribution and the proposed estimation technique. Through this data, we illustrate the applicability of the likelihood ratio test towards rejecting several well-known lifetime distributions that are nested within the wider class of EW distributions.
\end{abstract}

\section{Introduction}\label{sect1} 
{\hspace{7mm}}{\it Immune} or {\it cured} individuals in the context of survival analysis refers to subjects who would not encounter the event of interest under study, e.g., death due to a disease, relapse of a condition or return to prison (\citealp{maller1996survival}). Consequently, the observed lifetimes for the {\it immune} individuals  would always concur with the length of the study. Hence, the {\it immune} individuals would be indiscernible from the {\it censored} yet {\it non-immune} or {\it susceptible} individuals. \cite{maller1996survival} have pointed out that the presence of {\it cured} individuals in a survival data can be statistically tested. Ordinary survival analysis techniques ignore the presence of the fraction of individuals who are {\it cured}, commonly known as the cure fraction or cure rate. Therefore, several modified modeling techniques (known as the cure rate models) to analyze time to event data marked by the presence of cure fraction have been studied over the years (e.g., \citealp{kuk1992mixture, sy2000estimation, peng2000nonparametric, ibrahim2001bayesian, rodrigues2009poisson, balakrishnan2012algorithm, pal2018lbp}). Cure rate models have been applied extensively on cancer survival data for cancers with relatively better prognosis (e.g., melanoma, breast cancer, leukemia and prostate cancer), recidivism studies, and defaulting on a loan in credit risk assessment studies (\citealp{maller1996survival, de2017zero, ibrahim2014b}).\\

{\hspace{7mm}} The mixture cure rate model, also called the Bernoulli cure rate model, introduced by \cite{berkson1952survival} is probably the most widely used cure rate model. Under the mixture cure rate model, the overall population lifetime $Y$ is defined as 
\begin{equation}\label{eq1}
Y=\eta Y_s+(1-\eta)Y_c,
\end{equation}  
where $Y_s$ denotes the survival time for any {\it susceptible} individual, $Y_c = \infty$ denotes the survival time for any {\it cured} individual and $\eta$ is a random variable taking the value 1 or 0 depending on whether an individual is {\it susceptible} or {\it immune}, respectively. The model in (\ref{eq1}) can be further represented by
\begin{equation}\label{eq2}
S_p(y)=(1-\pi_0)S_s(y)+\pi_0,
\end{equation}    
where $S_p(.)$ and $S_s(.)$ are the respective survival functions corresponding to $Y$ and $Y_s$, and $\pi_0=P(\eta=0)$ is the cure rate. The mixture cure rate model has been explored in detail by several authors including \cite{farewell1982use}, \cite{goldman1984survivorship}, \cite{kuk1992mixture}, and \cite{sy2000estimation} with various assumptions and extensions. An alternative representation of the cure rate model, namely, the promotion time cure rate model was suggested by \cite{yakovlev1996stochastic} and was later investigated by \cite{chen1999new}, \cite{ibrahim2001bayesian}, \cite{yin2005cure}, \cite{ibrahim2014b}, and \cite{de2010note}, among others. Letting $M$ denote the random variable representing the number of competing causes, and $Y_j, j=1, \dots, M$, denote the promotion time or survival time corresponding to the $j$th cause, the overall population survival function $S_p(.)$ can be expressed as:
\begin{equation}\label{eq3}
S_p(y)=P(M=0)+ P(Y_1>y, \dots, Y_M>y|M\ge1 )P(M\ge 1)=\tilde g(S(y)),
\end{equation}    
where, given $M=m$, $Y_j, j=1, \dots, m$, are independently and identically distributed with a common survival function $S(.)$, and $\tilde g(.)$ is the probability generating function of $M$. Note that, in \eqref{eq3}, $M$ is unobserved, $Y_j, j=1, \dots, M,$ are independent of $M$, and $Y=\min\{Y_0, Y_1, \dots, Y_M\}$ is the actual lifetime of an individual with $P(Y_0=\infty)=1$. In cancer studies, competing causes may refer to the tumor cells that can potentially metastasize and cause detectable cancer. Several authors such as \cite{chen1999new} and  \cite{ibrahim2001bayesian} have assumed $M$ to follow a Poisson distribution, whereas \cite{rodrigues2009poisson}, \cite{balakrishnan2012algorithm}, \cite{pal2016likelihood}, \cite{balakrishnan2017proportional} and \cite{pal2017smmr} have modeled $M$ by a flexible Conway-Maxwell (COM) Poisson distribution. When $M$ is assumed to follow a Poisson distribution with mean $\theta$, $S_p(y)$ in \eqref{eq3} reduces to $S_p(y) = e^{-\theta F(y)}$ and the cure rate is given by $\pi_0=e^{-\theta}$.\\

{\hspace{7mm}} The survival function $S_s(y)$ in \eqref{eq2} or $S(y)$ in (\ref{eq3}) for any {\it susceptible} individual could be modeled and estimated by both parametric and non-parametric methods. From the statistical literature, positive valued continuous distributions like Weibull, gamma, generalized gamma and log normal distributions have been applied to model $Y_s$  or $Y_j$ (refer \citealp{farewell1982use, balakrishnan2013lognormal,  balakrishnan2014algorithm,  balakrishnan2015likelihood}). Semiparamteric generalizations to the model by assuming proportional hazards structure for $Y_s$ or  $Y_j$ have been discussed by \cite{kuk1992mixture}, \cite{chen1999new}, \cite{sy2000estimation}, and \cite{peng2000nonparametric}, whereas a class of semiparametric transformation models have been studied by \cite{yin2005cure}, \cite{li2002semi}, \cite{zeng2006semiparametric} and \cite{lu2004semiparametric}. Applications of piecewise constant and linear functions to estimate the baseline hazard function under proportional hazards model were discussed by \cite{ibrahim2001bayesian} and \cite{fotios}.         \\

{\hspace{7mm}} Missing data play an important role in the analysis of data with cure fraction where incompleteness in the data comes in two folds. Firstly, if censored, the information on the actual survival time of an individual is missing. Secondly, the information on the {\it cured} status is also missing for an individual who is censored. Therefore, parameter estimation may be challenging for the cure rate models. Several methods of estimating the model parameters as well as the baseline hazard or survival functions have been implemented, including ordinary maximum likelihood (ML) estimation (\citealp{farewell1982use}), Monte Carlo approximation of a marginal likelihood (\citealp{kuk1992mixture}), expectation-maximization (EM) algorithm (\citealp{sy2000estimation, chen2001maximum, peng2000nonparametric,balakrishnan2012algorithm}), profile likelihood, restricted non parametric ML estimation  (\citealp{tsodikov2003estimating}), unbiased estimating equations (\citealp{lu2004semiparametric, barui2020semiparametric}) and projected non-linear conjugate gradient technique based estimation (\citealp{pal2020new, pal2021new}). Very recently, \cite{davies2020stochastic} have introduced a stochastic version of the EM algorithm in the context of cure rate models where $Y_j$ is modeled by a generalized exponential distribution for every $j=1, \dots, M$.\\

{\hspace{7mm}} For the Bernoulli cure rate model, $M=0$ or $1$, and $Y=\min\{Y_0, Y_1\}$. In this manuscript, our main contribution is in the development of the stochastic expectation-maximization (SEM) algorithm to find the estimates of the parameters of the Bernoulli (mixture) cure rate model. In this regard, we propose to model the lifetime $Y_1$ by the flexible exponentiated Weibull (EW) distribution, which has not been studied before in the context of cure rate models. Being introduced by \cite{celeux1985sem}, the SEM algorithm has been designed to precisely estimate parameters in cases where the log-likelihood function has multiple stationary points, and the EM algorithm does not guarantee convergence to the significant local maxima. Unlike the EM algorithm, the SEM technique is less sensitive to the initial parameter choices, and the implementation is less cumbersome since it does not involve derivation of explicit expected values (\citealp{chauveau1995stochastic, celeux1996stochastic}). In particular, we show that the proposed SEM algorithm is more robust to the choice of initial values when compared to the EM algorithm.\\

{\hspace{7mm}}  The probability density function (pdf) of $Y_1$, under the assumption of EW distribution, is expressed as:
\begin{equation}\label{eq4}
f_s(y_1)=f_s(y_1; \alpha, k, \lambda)= \left(\frac{\alpha k}{\lambda}\right) \left(\frac{y_1}{\lambda}\right)^{k-1} e^{-(y_1/\lambda)^k}\left[1- e^{-(y_1/\lambda)^k}\right]^{\alpha-1},
\end{equation}    
where $y_1>0$ is the support of the distribution, $\alpha>0$ and $k >0$ are the shape parameters, and $\lambda>0$ denotes the scale parameter characterizing the distribution. The EW distribution has been introduced by \cite{mudholkar1993exponentiated} as an extension to the Weibull distribution by considering an additional shape parameter to the model. As pointed out by  \cite{mudholkar1996exponentiated} and \cite{khan2018exponentiated}, modeling failure times by an EW distribution is parsimonious as it accommodates both monotone increasing ($k\alpha \ge 1, k\ge1$) or decreasing ($k \alpha \le 1, k \le 1$), and non-monotone unimodal  ($k\alpha>1, k<1$) or bathtub shaped ($k\alpha<1, k>1$) hazard functions. Moreover, EW encompasses many well-known lifetime distributions as special cases, e.g., exponential ($\alpha=k=1$), Rayleigh ($\alpha=1, k=2$), Weibull ($\alpha=1$), generalized or exponentiated exponential ($k=1$), and Burr Type X ($k=2$) distributions. As a result, one can carry out hypotheses tests and model discrimination to validate if the sub models fit better. Furthermore, EW serves as an alternative to the generalized gamma distribution (\citealp{stacy1962generalization}), which is known to accommodate both monotone and non-monotone hazard functions. Interested readers can further refer to \cite{nassar2003exponentiated} and \cite{nadarajah2013exponentiated} for additional discussions on properties, applications and review on the EW distribution. \\

{\hspace{7mm}} The remainder of this manuscript is arranged in the following manner. We provide our model descriptions for the Bernoulli (mixture) cure rate model and basic properties of the EW distribution in Section \ref{sect2}. Section \ref{sect3} deals with the structure of the observed data and development of the likelihood function. In Section \ref{sect4}, we discuss the implementation of both the EM and SEM algorithms for estimating the model parameters and their standard errors. An extensive simulation study with carefully chosen parameter settings is carried out in Section \ref{sect5} to examine the robustness and accuracy of the estimation techniques. A model discrimination using likelihood-based criterion is performed to assess the flexibility of the EW distribution and the performance of the likelihood ratio test to correctly identify the true distribution. In Section \ref{sect6}, the flexibility of the proposed model and the performance of the estimation method are further substantiated based on real-life data collected from a malignant melanoma study. Finally, we provide some concluding remarks and scope of future research in Section \ref{sect7}.    
          
\section{Model descriptions}\label{sect2}
\subsection{Exponentiated Weibull lifetime distribution}
{\hspace{7mm}}  We assume the lifetime of the {\it susceptible} individuals to follow an EW distribution. Hence, the cumulative distribution function (cdf), survival function and hazard function of the {\it susceptible} lifetime $Y_1$ have the following forms:
\begin{equation}\label{eq5}
 F_s(y_1) =F_s(y_1; \alpha, k, \lambda)= \left[1-e^{-(y_1/\lambda)^k}\right]^{\alpha},
\end{equation}
\begin{equation}\label{eq6}
 S_s(y_1) =S_s(y_1; \alpha, k, \lambda)= 1- \left[1-e^{-(y_1/\lambda)^k}\right]^{\alpha},
\end{equation}
and 
\begin{equation}\label{eq7}
h_s(y_1) = h_s(y_1; \alpha, k, \lambda)=\frac{\left(\frac{\alpha k}{\lambda}\right) \left(\frac{y_1}{\lambda}\right)^{k-1} e^{-(y_1/\lambda)^k}\left[1- e^{-(y_1/\lambda)^k}\right]^{\alpha-1}}{1-\left[1-e^{-(y_1/\lambda)^k}\right]^{\alpha}},
\end{equation}
respectively, where $y_1>0$, $\alpha>0, k>0$ and $\lambda>0$. One interesting interpretation of the EW distribution is in the area of reliability. If there are $n$ components in a parallel system and the lifetimes of the components are independently and identically distributed as EW, then the system lifetime also follows an EW distribution. As pointed out by \cite{nadarajah2013exponentiated}, EW finds applications in a wide variety of problems, e.g., modeling extreme value data on water discharge arising due to river floods, data on optimal accelerated life test plans under type I censoring, firmware system failures, software release times, fracture toughness of materials, bus motor failures  and number of ozone peaks, among others.  
From \cite{mudholkar1993exponentiated} and \cite{mudholkar1996exponentiated},  we note that
\begin{enumerate}
    \item [(a)] if $\alpha=k=1$, then the hazard rate is constant;
\item [(b)] if $\alpha=1$, the hazard rate is increasing for $k>1$ and decreasing for $k<1$;
\item [(c)] if $k=1$, the hazard rate is increasing for $\alpha>1$ and decreasing for $\alpha<1$.
\end{enumerate}
Additionally, the combinations of the two shape parameters as presented in Table \ref{Tab1} render various shapes to the hazard function.
\begin{table}[h!]
    \centering
   \caption{Hazard rate pattern across various parameter values}   
   \begin{tabular}{ccc}\hline
    $k$  &$\alpha$ &hazard-rate \\\hline
1& $< 1 $& constant (exponential)\\
&$> 1$  &monotonic (Weibull)\\
$< 1$& $< 1$ &decreasing\\
$> 1$&$>1$ &increasing\\
$>1$ &$< 1$&bathtub or increasing\\ 
$<1$&$> 1$&unimodal or decreasing\\\hline
    \end{tabular}
    \label{Tab1}
\end{table}
The general expression for  the $q$th order raw moment for a random variable $Y_1$ following the EW distribution has been derived by  \cite{pal2006exponentiated}, which is given by 
\begin{equation}\label{eq8}
    E(Y_1^q)=\begin{cases}\alpha \lambda^q \Gamma\left(\frac{q}{k}+1\right)\sum_{i=0}^{\alpha-1} {\alpha-1 \choose i} (-1)^i (i+1)^{-\frac{q}{k}-1}, & \text{if $\alpha \in N$}, \\
    \alpha \lambda^q \Gamma\left(\frac{q}{k}+1\right)\sum_{i=0}^{\infty}\frac{(\alpha-1)(\alpha-2)\dots (\alpha-i)}{i!} (-1)^i (i+1)^{-\frac{q}{k}-1}, & \text{if $\alpha \notin N$, for $q=0,1,2,..,$}
    \end{cases}
\end{equation}
where $N$ denotes the set of natural numbers.


\subsection{Bernoulli (mixture) cure rate model}
{\hspace{7mm}} On assuming the number of competing causes $M$ to follow a Bernoulli distribution, i.e., there is either a single cause that can result in an event of interest or there is no cause resulting in a cure, the probability mass function (pmf) of $M$ can be expressed as:
\begin{equation}\label{eq8.1}
p(m; \nu)=P(M=m; \nu)= \left(\frac{1}{1+\nu}\right)^{1-m}\left(\frac{\nu}{1+\nu}\right)^{m}, m=0, 1,
\end{equation}
where $\nu>0$. The survival function of the random variable $Y=\min\{Y_0,Y_1\}$, also referred to as the population survival function, can be obtained by combining (\ref{eq3}) and (\ref{eq6}) and is given by
\begin{flalign}\label{eq8.2}
S_p(y)&=S_p(y; \nu, \alpha, k, \lambda) = P(Y>y)\nonumber \\ &=\sum_{m=0}^1 P\left(Y> y|M=m\right)P(M=m; \nu)\nonumber\\
&=p(0; \nu) +  S_s(y)p(1; \nu) \nonumber\\
&= \left(\frac{1}{1+\nu}\right)+\left\{ 1- \left[1-e^{-(y/\lambda)^k}\right]^{\alpha}\right\}\left(\frac{\nu}{1+\nu}\right).
\end{flalign}
Further, note that 
\begin{flalign}\label{eq8.3}
 \underset{y \to \infty}{\lim}  S_p(y) = \left(\frac{1}{1+\nu}\right) = \pi_0
\end{flalign}
is the cure rate or cure probability of any individual in the population. Hence, the population density function can be derived from (\ref{eq8.2}) as: 
\begin{flalign}\label{eq8.4}
f_p(y)&=f_p(y; \nu, \alpha, k, \lambda) = -\frac{dS_p(y)}{dy} = \left(\frac{\nu}{1+\nu}\right) \left(\frac{\alpha k y^{k-1}}{\lambda^k}\right)  e^{-(y/\lambda)^k}\left[1- e^{-(y/\lambda)^k}\right]^{\alpha-1}.
\end{flalign}


\section{Form of the data and likelihood function}\label{sect3}
{\hspace{7mm}}  The right censoring scheme is considered in our study. For $i=1, \dots, n$ with $n$ denoting the sample size, let $Y_i$ and $C_i$ respectively denote actual survival time and censoring time for the individual $i$. Let $\delta_i=I(Y_i \le C_i)$ be the censoring indicator and $T_i=\min\{Y_i, C_i\}$ be the observed lifetime for the $i$th individual. 
Therefore, the observed survival data is represented in the form of a triplet denoted by $\{(t_i, \delta_i, \bm x^*_i): i=1, \dots, n\}$ where $t_i$ is a realization of $T_i$ and $\bm x_i^{*}=(x_{1i}, \dots, x_{di})^{\tiny \rm T} \in \mathbb{R}^d$  is the $d$-dimensional covariate vector specific to the $i$th subject. Let $\bm x_i=\left(1, \bm x_i^{*\tiny \rm T}\right)^{\tiny \rm T} \in \mathbb{R}^{d+1}$. We further denote $\bm X=\left(\bm x_1, \dots, \bm x_n \right)^{\tiny \rm T} \in \mathbb{R}^{(d+1) \times n}$, $\bm \delta=(\delta_1, \dots, \delta_n)^{\tiny \rm T} \in \mathbb{B_{\delta}}$ and $\bm t=(t_1, \dots, t_n)^{\tiny \rm T} \in \mathbb{R}^{n}_{>0}$, where $\mathbb{B_{\delta}}=\{(\delta_1, \dots, \delta_n): \delta_i=0, 1 \text{ }\forall i=1, \dots, n\}$. In order to associate the effect of covariates to the cure rate for every $i=1, \dots, n$, we use log-linear function $\nu_i=e^{\bm x_i^{\tiny \rm T} \bm \beta}$ to link the parameter $\nu>0$ with the covariate vector $\bm x_i$, where $\bm \beta=(\beta_0, \beta_1, \dots, \beta_d)^{\tiny \rm T}$ is the respective $(d+1)$-dimensional vector of regression parameters. \\

{\hspace{7mm}} We define $\bm \theta=\left(\bm \beta^{\tiny \rm T}, \alpha, k, \lambda \right)^{\tiny \rm T} \in \bm \Theta \subset \mathbb{R}^{d+4}$ as the unknown parameter vector and $\bm \Theta$ as the parameter space. Therefore, the likelihood function $L_O(\bm \theta; \bm t, \bm \delta, \bm X)$ based on the observed data is given by 
\begin{flalign}\label{eq9}
L_O(\bm \theta; \bm t, \bm \delta, \bm X) \propto \prod_{i=1}^n \{f_p(t_i; \bm \theta, \delta_i, \bm x^*_i)\}^{\delta_i} \{S_p(t_i; \bm \theta, \delta_i, \bm x^*_i)\}^{1-\delta_i},
\end{flalign}  
where  $S_p(.; \bm \theta, \delta_i, \bm x^*_i)$ and $f_p(.; \bm \theta, \delta_i, \bm x^*_i)$ denote the respective population density and survival functions for individual $i$, and can be obtained from (\ref{eq8.2}) and (\ref{eq8.4}) respectively with some notation adjustments. Hence, the observed data log-likelihood function is expressed as:
\begin{flalign}\label{eq10}
l_O(\bm \theta; \bm t, \bm \delta, \bm X) & =\log L_O(\bm \theta; \bm t, \bm \delta, \bm X) \nonumber \\ 
& =\text{constant} + \sum_{i=1}^n \delta_i \log f_p(t_i; \bm \theta, \delta_i, \bm x^*_i) + \sum_{i=1}^n (1-\delta_i) \log S_p(t_i; \bm \theta, \delta_i, \bm x^*_i).
\end{flalign}  
Let us define $\Delta_1=\{i: \delta_i=1\}, \Delta_0=\{i: \delta_i=0\}, n_1=|\Delta_1|$ and $F_w(t_i; k, \lambda)=1-e^{-(t_i/\lambda)^k}$ for $i=1, \dots, n$. \\

{\hspace{7mm}} From (\ref{eq8.2}), (\ref{eq8.4}), (\ref{eq10}) and using $\nu_i=e^{\bm x_i^{\tiny \rm T} \bm \beta}, i=1, \dots, n$, the log-likelihood function for the Bernoulli cure rate model takes the following form:
\begin{flalign}\label{eq11}
l_O(\bm \theta; \bm t, \bm \delta, \bm X) &=\text{constant} + \sum_{i=1}^n \delta_i \log \left[\frac{e^{\bm x_i^{\tiny \rm T}\bm \beta}}{1+e^{\bm x_i^{\tiny \rm T}\bm \beta}}   \left(\frac{\alpha k t_i^{k-1}}{\lambda^k}\right)  e^{-(t_i/\lambda)^k}\left\{1- e^{-(t_i/\lambda)^k}\right\}^{\alpha-1}
\right] \nonumber \\
&+ \sum_{i=1}^n (1-\delta_i) \log \left[\frac{1+e^{\bm x_i^{\tiny \rm T}\bm \beta}\left\{1-\left[1-e^{-(t_i/\lambda)^k}\right]^{\alpha}\right\}}{1+e^{\bm x_i^{\tiny \rm T}\bm \beta}}\right] \nonumber\\
&=\text{constant}+ n_1(\log \alpha+\log k - k\log \lambda)\nonumber \\
&+ \sum_{i \in \Delta_1} \left\{\bm x_i^{\tiny \rm T}\bm \beta - \log \left(1+e^{\bm x_i^{\tiny \rm T}\bm \beta}\right)+(k-1)\log t_i - \left(\frac{t_i}{\lambda}\right)^k + (\alpha-1) \log F_w(t_i; k, \lambda)\right\}\nonumber\\
&+ \sum_{i \in \Delta_0}\left[ \log\left\{1+e^{\bm x_i^{\tiny \rm T}\bm \beta}\left[1-F_w(t_i; k, \lambda)^{\alpha}\right]\right\} - \log \left(1+e^{\bm x_i^{\tiny \rm T}\bm \beta}\right) \right].\\
\end{flalign}  

{\hspace{7mm}} The expressions of the first order and second order derivatives of $l_O(\bm \theta; \bm t, \bm \delta, \bm X)$ with respect to $\bm \theta$ are presented in the Supplemental Material. These expressions would allow interested researchers to directly maximize $l_O(\bm \theta; \bm t, \bm \delta, \bm X)$ to obtain an estimate of $\bm \theta$. However, the presence of missing data (due to censoring) strongly motivates us to develop algorithms that can handle such missingness of data.


\section{Estimation techniques}\label{sect4}
{\hspace{7mm}} As defined in Section \ref{sect1}, let $\eta_i=1$ if an individual is not {\it cured} and $\eta_i=0$ is an individual is {\it cured}, for $i=1, \dots, n$. It can be seen that $\eta_i=1$ for $i \in \Delta_1$ and $\eta_i$ is unknown (hence, is missing) for $i \in \Delta_0$. The data we observe is partial, and hence, the problem can be treated as an incomplete data problem. Therefore, the EM or EM like algorithms can be applied for the ML or ML type estimation of $\bm \theta$. 
\subsection{Expectation maximization (EM) algorithm}\label{sect4.1}
 
{\hspace{7mm}} Introduced by \cite{dempster1977maximum}, the EM algorithm is a popular and well accepted iterative technique of obtaining ML estimates based on incomplete data. The popularity is legitimate since the algorithm is easy to implement and ensures monotonicity of the likelihood function towards the local maxima. To implement the EM algorithm, we define the complete data likelihood function as:
\begin{equation}\label{eq12}
L_C(\bm \theta; \bm t, \bm \delta, \bm X, \bm \eta) \propto \prod_{i \in \Delta_1} f_p(t_i; \bm \theta, \delta_i, \bm x^*_i) \times \prod_{i \in \Delta_0}\left\{\pi_0(\bm x^*_i; \bm \beta)\right\}^{1-\eta_i}\left\{ S_p(t_i; \bm \theta, \delta_i, \bm x^*_i) - \pi_0(\bm x^*_i; \bm \beta) \right\}^{\eta_i},
\end{equation}
where $\bm \eta=\left(\eta_1, \dots, \eta_n\right)^{\tiny \rm T}$ and $\pi_0(\bm x^*_i; \bm \beta)=\left\{1+e^{\bm x_i^{\tiny \rm T} \bm \beta}\right\}^{-1}$ is the cure rate. Equivalently, the expression for the complete data log-likelihood function is obtained as:
\begin{flalign}\label{eq13}
l_C(\bm \theta; \bm t, \bm \delta, \bm X, \bm \eta)& = \text{constant} + \sum_{i \in \Delta_1} \log f_p(t_i; \bm \theta, \delta_i, \bm x^*_i) + \sum_{i \in \Delta_0}(1-\eta_i) \log \pi_0(\bm x^*_i; \bm \beta) \nonumber\\& +\sum_{i \in \Delta_0}{\eta_i}\log \left\{ S_p(t_i; \bm \theta, \delta_i, \bm x^*_i) - \pi_0(\bm x^*_i; \bm \beta) \right\}.
\end{flalign}
For the Bernoulli cure rate model, the expression given in (\ref{eq13}) takes the following form:
\begin{flalign}\label{eq14}
l_{CB}(\bm \theta; \bm t, \bm \delta, \bm X, \bm \eta) & = \text{constant}+ n_1 (\log \alpha +\log k - k \log \lambda) + (k-1)\sum_{i \in \Delta_1} \log t_i \nonumber \\  &- \sum_{i \in \Delta_1} \left(\frac{t_i}{\lambda}\right)^k +  \sum_{i \in \Delta_1} (\alpha-1) \log \left\{ 1 - e^{-(t_i/\lambda)^k}\right\} + \sum_{i \in \Delta_1} \bm x_i^{\tiny \rm T} \bm \beta - \sum_{i=1}^n \log \left(1+ e^{\bm x_i^{\tiny \rm T} \bm \beta} \right)   \nonumber \\  & + \sum_{i \in \Delta_0} \eta_i e^{\bm x_i^{\tiny \rm T} \bm \beta} + \sum_{i \in \Delta_0} \eta_i \log\left\{ 1- \left[1 - e^{-(t_i/\lambda)^k}\right]^{\alpha}  \right\}.
\end{flalign}

\subsubsection*{Steps involved in the EM algorithm:} 
{\hspace{7mm}}Begin the iterative process by considering an initial estimate $\bm \theta^{(0)}=\left(\bm \beta^{(0)}, \alpha^{(0)}, k^{(0)}, \lambda^{(0)}\right)^{\tiny \rm T}$ of $\bm \theta$. The choice of  $\bm \theta^{(0)}$ requires justifications based on background knowledge and some sample real-life data. For $r=1, 2, \dots $, let $\bm \theta^{(r)}$ be the estimate of $\bm \theta$ at the $r$th step of the iteration. Then,  $\bm \theta^{(r+1)}$ is obtained using the following steps:
\begin{enumerate}
\item {\it E-Step}: Find the conditional expectation $Q\left(\bm \theta; \bm \theta^{(r)} \right)= E\left\{l_{CB}(\bm \theta; \bm t, \bm \delta, \bm X, \bm \eta)| \left(\bm \theta^{(r)}, \bm t, \bm \delta, \bm X \right)\right\}.$ As discussed in \cite{yang2016stochastic}, and using (\ref{eq1}) and (\ref{eq14}), we obtain 
\begin{flalign}\label{eq15}
Q\left(\bm \theta; \bm \theta^{(r)} \right) & = \text{constant}+ n_1 (\log \alpha +\log k - k \log \lambda) + (k-1)\sum_{i \in \Delta_1} \log t_i \nonumber \\  &- \sum_{i \in \Delta_1} \left(\frac{t_i}{\lambda}\right)^k +  \sum_{i \in \Delta_1} (\alpha-1) \log \left\{ 1 - e^{-(t_i/\lambda)^k}\right\} + \sum_{i \in \Delta_1} \bm x_i^{\tiny \rm T} \bm \beta - \sum_{i=1}^n \log \left(1+ e^{\bm x_i^{\tiny \rm T} \bm \beta} \right)   \nonumber \\  & + \sum_{i \in \Delta_0} E\left\{\eta_i \big \vert \left(\bm \theta^{(r)}, \bm t, \bm \delta, \bm X \right)\right\} e^{\bm x_i^{\tiny \rm T} \bm \beta} + \sum_{i \in \Delta_0} E\left\{\eta_i \big \vert \left(\bm \theta^{(r)}, \bm t, \bm \delta, \bm X \right)\right\} \log\left\{ 1- \left[1 - e^{-(t_i/\lambda)^k}\right]^{\alpha}  \right\},
\end{flalign}
where 
\begin{flalign}\label{eq16}
E\left\{\eta_i \big \vert \left(\bm \theta^{(r)}, \bm t, \bm \delta, \bm X \right)\right\} &= P\left\{\eta_i  = 1\big \vert \left(\bm \theta^{(r)}, \bm t, \bm \delta, \bm X \right)\right\}\nonumber \\
&=P\left\{\eta_i = 1\big \vert \left(\bm \theta^{(r)}, Y_i>t_i, \bm x^*_i, i \in \Delta_0   \right)\right\} \nonumber \\
&=\frac{P\left\{Y_i>t_i\big \vert \left( \eta_i=1, \bm \theta^{(r)},  \bm x^*_i, i \in \Delta_0   \right) \right\}P\left\{\eta_i=1 \big \vert \left(\bm \theta^{(r)},  \bm x^*_i, i \in \Delta_0\right) \right\}}{P\left\{Y_i>t_i\big \vert \left( \bm \theta^{(r)},  \bm x^*_i, i \in \Delta_0   \right) \right\}} \nonumber \\
&=\frac{S_p\left(t_i; \bm \theta^{(r)}, \delta_i, \bm x^*_i\right) - \pi_0\left(\bm x^*_i; \bm \beta^{(r)}\right)}{S_p\left(t_i; \bm \theta^{(r)}, \delta_i, \bm x^*_i\right)}\nonumber \\
&=1- \frac{ \pi_0\left(\bm x^*_i; \bm \beta^{(r)}\right)}{S_p\left(t_i; \bm \theta^{(r)}, \delta_i, \bm x^*_i\right)}.
\end{flalign}
\item {\it M-Step}: Find 
\begin{equation}\label{eq17}
\bm \theta^{(r+1)}=\left(\bm \beta^{(r+1)}, \alpha^{(r+1)}, k^{(r+1)}, \lambda^{(r+1)}\right)^{\tiny \rm T} = \underset{\bm \theta}{{\arg \max}}    
\text{ }Q\left(\bm \theta; \bm \theta^{(r)} \right).
\end{equation}      
The maximization step can be carried out using multidimensional unconstrained optimization methods like Nelder-Mead simplex search algorithm or quasi Newton methods like BFGS algorithm (see \citealp{fletcher2013practical}). These algorithms are available in statistical software R version 4.0.3 under General Purpose Optimization package called \texttt{optimr()}.  
\item {\it Convergence}: Check if the stopping or convergence criterion for the iterative process is met. For our analysis,  we consider that the EM algorithm has converged to a local maxima if
\begin{equation}\label{eq18}
\underset{1 \le k' \le d+4}{{\max}}\text{ }{\left \vert  \frac{\theta^{(r+1)}_{k'}- \theta^{(r)}_{k'}}{\theta^{(r)}_{k'}} \right \vert < \epsilon },
\end{equation} 
where $\theta^{(r)}_{k'}$ and $\theta^{(r+1)}_{k'}$ are the $k'$th component of $\bm \theta^{(r)}$ and $\bm \theta^{(r+1)}$, respectively, and $\epsilon$ is a tolerance such as 0.001.
\end{enumerate}  
If the condition in $(\ref{eq18})$ is satisfied, then the iterative process is stopped and  $\bm \theta^{(r)}$ is considered as the ML estimate of $\bm  \theta$ (\citealp{mclachlan2007algorithm, pal2017ieee,pal2017cs}).

\subsection{Stochastic expectation maximization (SEM) algorithm}\label{sect4.2}  

{\hspace{7mm}} As discussed in  \cite{dempster1977maximum} and \cite{mclachlan2007algorithm}, the sequence $\left\{\bm \theta^{(r)}\right\}$, as obtained by implementing the EM algorithm, gradually converges to a stationary point of $L_{O}(\bm \theta; \bm t, \bm \delta, \bm X)$. However, the convergence rate depends on factors such as choice of initials parameter values and the flatness of likelihood surface. Further, for likelihood surfaces characterized by several stationary points including saddle points, the EM algorithm does not guarantee convergence to the significant local maxima. It is also noted that the rate of convergence of the EM algorithm is heavily influenced by the proportion of missing observations. Moreover, analytical steps in deriving conditional expectation involve computation of integrals which is often intensive, complex, and in some cases, intractable. In our considered modeling framework, and as we have seen, computation of the conditional expectations is not complicated. However, the EM may be quite sensitive to the choice of initial values, which motivates the development of an alternate algorithm. \\

{\hspace{7mm}}To address the issues related to the EM algorithm, the SEM algorithm works on the idea of simulating pseudo values to replace the missing values. The SEM comprises two steps, namely, the S-step and the M-step.  The S-step involves generating a pseudo sample from the conditional distribution of the missing data given the observed information and current parameter values. The M-step involves finding the parameter value which maximizes the complete data log-likelihood function based on the pseudo sample (\citealp{celeux1985sem, celeux1992stochastic, celeux1996stochastic}). The random generation of values to impute missing data allows the SEM algorithm to overcome the problem of getting trapped in an insignificant local maxima or saddle point (\citealp{celeux1996stochastic, bordes2007stochastic, cariou2008unsupervised}). A discussion on the asymptotic properties based on a mixture model reveals that the sequence of estimates generated by the SEM algorithm converges to a stationary Gaussian distribution whose mean is the consistent ML estimator of the mixing proportion (\citealp{diebolt1993asymptotic}).  In their paper, \cite{svensson2010asymptotic} and \cite{cariou2008unsupervised} established that SEM works well for relatively smaller sample sizes and the algorithm is less sensitive to  initial parameter choices. \\

{\hspace{7mm}}Define $\tilde H_1=\{i: \eta_i=1\}$ and $\tilde H_0=\{i: \eta_i=0\}$. Note that $\tilde H_0$ is unobserved and $\tilde H_1$ is only partially observed. Hypothetically, assuming that we completely observe $\tilde H_0$ and $\tilde H_1$, then, for any individual $i \in \tilde H_0$, $Y_i>C_i$ and the contribution by $i$ to the likelihood function would be through the cure rate $\pi_0(\bm x^*_i; \bm \beta)$. Again, for any $i \in \tilde H_1$, $Y_i > C_i$ or $Y_i \le C_i$, $T_i=\min\{Y_i, C_i\}$ and contribution to the likelihood function by $i$ would be through the population density function $f_p(t_i; \bm \theta, \delta_i, \bm x^*_i)$. For the latter, the information on the actual lifetime is missing if the individual is right censored, and observed when not censored. Therefore, we would stochastically generate both cured status $\eta_i$ and subject's actual lifetime $y_i^*$, and hence, generate pseudo data of the form $\left\{(y^*_i, \delta_i, \bm x^*_i, \eta_i): i = 1, \dots, n\right\}$. \\     

{\hspace{7mm}}To implement the SEM algorithm, unlike the EM algorithm, the complete data likelihood and log-likelihood functions are defined by 
\begin{equation}\label{eq19}
\tilde L_C(\bm \theta; \bm y^*, \bm \delta, \bm X, \bm \eta) \propto \prod_{i=1}^n  \left\{f_p(y^*_i; \bm \theta, \delta_i, \bm x^*_i)\right\}^{\eta_i}  \left\{\pi_0(\bm x^*_i; \bm \beta)\right\}^{1-\eta_i}
\end{equation}
and 
\begin{equation}\label{eq20}
\tilde l_C(\bm \theta; \bm y^*, \bm \delta, \bm X, \bm \eta) = \text{constant}+\sum_{i=1}^n \eta_i \log \left\{f_p(y^*_i; \bm \theta, \delta_i, \bm x^*_i)\right\}  + (1-\eta_i) \log \left\{\pi_0(\bm x^*_i; \bm \beta)\right\},
\end{equation}
respectively, where $y_i^*$ denotes the actual lifetime generated stochastically for $i \in \Delta_0$, and $\bm y^*=\left(y_1^*, \dots, y^*_n\right)^{\tiny \rm T}. $ For the Bernoulli cure rate model, (\ref{eq20}) becomes
\begin{flalign}\label{eq21}
\tilde l_{CB}(\bm \theta; \bm y^*, \bm \delta, \bm X, \bm \eta) &=  \text{constant}+  (\log \alpha +\log k - k \log \lambda) \sum_{i=1}^n \eta_i + (k-1)\sum_{i=1}^n \eta_i \log y^*_i \nonumber \\  &- \sum_{i=1}^n \eta_i \left(\frac{y^*_i}{\lambda}\right)^k +  (\alpha-1) \sum_{i=1}^n \eta_i \log \left\{ 1 - e^{-(y^*_i/\lambda)^k}\right\} - \sum_{i=1}^n (1-\eta_i)\log \left(1+ e^{\bm x_i^{\tiny \rm T} \bm \beta} \right).
\end{flalign}

\subsubsection*{Steps involved in the SEM algorithm:} 
{\hspace{7mm}}Similar to the EM algorithm implementation, start the iterative process for the SEM algorithm with a reasonable initial choice $\bm \theta^{(0)}=\left(\bm \beta^{(0)}, \alpha^{(0)}, k^{(0)}, \lambda^{(0)}\right)^{\tiny \rm T}$ of the parameter $\bm \theta$. For some pre-defined $R \in \mathbb{Z}^+$ and $r=0, 1, \dots, R$, assume $\bm \theta^{(r)}=\left(\bm \beta^{(r)}, \alpha^{(r)}, k^{(r)}, \lambda^{(r)}\right)^{\tiny \rm T}$ as the estimate of the parameter $\bm \theta$ for the $r$th step. The steps below permit the computation of the ML type estimate of $\bm \theta$ by applying the SEM algorithm.

\begin{enumerate}
\item {\it S-Step}: There are two sub-steps to be followed in the stochastic step of the implementation.  
\begin{enumerate}
\item [A.] {\it Generating cure status $\eta_i^{(r+1)}$ for $i=1, \dots, n$}:
\begin{itemize}
 \item [(i)] For $i \in \Delta_1$, $\eta^{(r+1)}_i=1$. 
\item [(ii)] For $i \in \Delta_0$, generate $\eta_i^{(r+1)}$ from a Bernoulli distribution with conditional probability of success $p^{(r+1)}_{s,i}$ using (\ref{eq16}) as: 
\begin{flalign}\label{eq22}
p^{(r+1)}_{s,i} = P\left\{\eta_i^{(r+1)}=1 \Big\vert \left( \bm \theta^{(r)}, Y_i>t_i, \bm x^*_i, i \in \Delta_0 \right)\right\} = 1- \frac{ \pi_0\left(\bm x^*_i; \bm \beta^{(r)}\right)}{S_p\left(t_i; \bm \theta^{(r)}, \delta_i, \bm x^*_i\right)}.
\end{flalign}  
\end{itemize}
\item [B.] {\it Generating actual lifetime $y^{*(r+1)}_i$ for $i=1, \dots, n$}:
\begin{itemize}
\item [(i)] For $i \in \Delta_1$, $y^{*(r+1)}_i = t_i$ is the actual lifetime.
\item [(ii)] For $i \in \Delta_0$ and if $\eta^{(r+1)}_i=0$ from step  1A., $y^{*(r+1)}_i=\infty$ since the individual is {\it cured} with respect to the event of interest.  
\item [(iii)] For $i \in \Delta_0$ and if $\eta^{(r+1)}_i=1$ from 1A., we only observe the censoring time $t_i=c_i$ since the actual lifetime $Y_i>t_i$. Hence, actual lifetime $y^{*(r+1)}_i$ is generated from a truncated EW distribution with density $g(.; \bm \theta^{(r)}, \delta_i, \bm x^*_i)$, where 
\begin{equation}\label{eq23}
g\left(y^{*(r+1)}_i; \bm \theta^{(r)}, \delta_i, \bm x^*_i\right)
=\frac{f_p\left(y^{*(r+1)}_i; \bm \theta^{(r)}, \delta_i, \bm x^*_i\right)}{S_p\left(t_i; \bm \theta^{(r)}, \delta_i, \bm x^*_i\right)}, \ \ y^{*(r+1)}_i>t_i \text{ with } \delta_i=0.
\end{equation} 
Let  $G(.; \bm \theta^{(r)}, \delta_i, \bm x^*_i)$ denote the cdf corresponding to $g(.; \bm \theta^{(r)}, \delta_i, \bm x^*_i)$. It can be noted that $G(.; \bm \theta^{(r)}, \delta_i, \bm x^*_i)$ is not a proper cdf as
\begin{flalign}\label{eq23.1}
G\left(y^{*(r+1)}_i; \bm \theta^{(r)}, \delta_i, \bm x^*_i\right) = 1- \frac{S_p\left(y^{*(r+1)}_i; \bm \theta^{(r)}, \delta_i, \bm x^*_i\right)}{S_p\left(t_i; \bm \theta^{(r)}, \delta_i, \bm x^*_i\right)}
\end{flalign} 
and 
\begin{flalign}\label{eq23.2}
\underset{ y_i^{*(r+1)}\to \infty}{{\lim}} G\left(y^{*(r+1)}_i; \bm \theta^{(r)}, \delta_i, \bm x^*_i\right) = 1- \frac{\pi_0\left(\bm x^*_i; \bm \beta^{(r)}\right)}{S_p\left(t_i; \bm \theta^{(r)}, \delta_i, \bm x^*_i\right)} = b_i^{(r+1)},
\end{flalign} 
where $ b_i^{(r+1)}=1$ only if $\pi_0\left(\bm x^*_i; \bm \beta^{(r)}\right)=0$. In this case, two schemes could be followed for generating $y^{*(r+1)}_i$.
\begin{itemize}
\item [(a)] Generate $u_i^{(r+1)}$ randomly from $\text{uniform}\left(0, b_i^{(r+1)}\right)$ and take an inverse transformation to find $y^{*(r+1)}_i=G^{-1}\left(u_i^{(r+1)}; \bm \theta^{(r)}, \delta_i, \bm x^*_i\right)$.  
\item [(b)] Generate $m_i^{(r+1)}$ using $p\left(m_i^{(r+1)}; e^{\bm x_i^{\tiny \rm T}\bm \beta^{(r)}}\right)$ given in (\ref{eq8.1}), i.e., from a Bernoulli distribution with success probability $\left\{\frac{e^{\bm x_i^{\tiny \rm T}\bm \beta^{(r)}}}{1+e^{\bm x_i^{\tiny \rm T}\bm \beta^{(r)}}}\right\}$. If $m_i^{(r+1)}=1$, then simulate $y_i^{*(r+1)}$ from the pdf $g(.; \bm \theta^{(r)}, \delta_i, \bm x^*_i)$, which is the pdf of a truncated EW distribution given in (\ref{eq23}), truncated at $t_i$.     
\end{itemize}
\end{itemize}
\end{enumerate}
\item {\it M-Step}: Once the pseudo complete data $\left\{\left(y^{*(r+1)}_i, \delta_i, \bm x^*_i, \eta_i^{(r+1)}\right): i = 1, \dots, n\right\}$ is obtained, find the updated estimate by
\begin{equation}\label{eq24}
\bm \theta^{(r+1)}=\left(\bm \beta^{(r+1)}, \alpha^{(r+1)}, k^{(r+1)}, \lambda^{(r+1)}\right)^{\tiny \rm T} = \underset{\bm \theta}{{\arg \max}}    
\text{ } \tilde l_{CB}(\bm \theta; \bm y^{*(r+1)}, \bm \delta, \bm X, \bm \eta^{(r+1)}),
\end{equation}      
where $\bm y^{*(r)}=\left(y_1^{*(r)}, \dots, y^{*(r)}_n\right)^{\tiny \rm T}$ and $\bm \eta^{*(r)}=\left(\eta_1^{*(r)}, \dots, \eta^{*(r)}_n\right)^{\tiny \rm T}$. The implementation of the {\it M-Step} follows the same techniques and routines as given in the {\it M-Step} of the EM algorithm.
\item Repeat steps 1 and 2 $R$ times to obtain the sequence of estimates $\left\{\bm \theta^{(r)}\right\}_{r=1}^R$. As pointed out by \cite{diebolt1993asymptotic}, the sequence $\underset{R \to \infty}{{\lim}} \left\{\bm \theta^{(r)}\right\}_{r=1}^R$  does not converge pointwise, and hence, the implementation of the SEM algorithm will not result in the consistent ML estimator. However, the ergodic Markov chain $\left\{\bm \theta^{(r)}\right\}_{r=1}^R$ generated by the implementation of the SEM algorithm converges to a normal distribution. It was further established by \cite{diebolt1993asymptotic} that the mean of the normal distribution is the consistent ML estimate of $\bm \theta$ under some mild technical assumptions. Based on this result and arguments provided by \cite{celeux1996stochastic} and \cite{davies2020stochastic}, the SEM estimate $\hat {\bm \theta}_{SEM}$ can be obtained by the following two approaches:
\begin{enumerate}
\item Calculate the SEM estimate  by   
\begin{equation}\label{eq25}
\hat {\bm \theta}_{SEM}= \{R-R^*\}^{-1} \sum_{r=R^*+1}^R \bm \theta^{(r)},
\end{equation}      
where iterations $r=1, \dots, R^*$ represent `burn-in' or `warm-up' period to reach the stationary regime, and the estimates $\bm \theta^{(r)}, r=1, \dots, R^*$ are discarded. \cite{marschner2001miscellanea} indicated that a point estimate of $\bm \theta$ can be calculated by taking average over the estimates obtained from iterations of the SEM algorithm after sufficiently long burn-in period. Both \cite{marschner2001miscellanea} and \cite{ye2014analysis} used first 100 iterations of the algorithm as the burn-in period, and considered additional 900 - 1000 iterations for obtaining the SEM estimates. However, it is recommended to do a trace plot of the sequence of estimates against iteration numbers to examine the trend in the behavior of the estimates, and thereby, choosing an appropriate burn-in period.      
\item Carry out $R^*$ iterations as `warm-up' and derive the sequence $\left\{\bm \theta^{(r)}\right\}_{r=1}^{R^*}$ by implementing the SEM algorithm. Find 
\begin{equation}\label{eq24}
\hat{\bm \theta}_{SEMI}= \underset{\{\bm \theta^{(r)}, r=1, \dots, R^*\}}{{\arg \max}} \text{ } l_{O}\left(\bm \theta; \bm t, \bm \delta, \bm X\right).
\end{equation}      
By taking  $\hat{\bm \theta}_{SEMI}$ as the starting value, the EM algorithm is implemented to derive the ML estimate $\hat {\bm \theta}$ (see  \citealp{celeux1992stochastic, celeux1996stochastic}).     \\
\end{enumerate}
\end{enumerate}

{\hspace{7mm}}Note that approach (b) above requires the development of both SEM and EM algorithms and hence may not be a preferred approach to calculate the estimates. On the other hand, approach (a) above may result in under-estimation of the variances of the estimators, see \cite{Dieb95}. In fact, in our model fitting study, as presented in Section \ref{sect5.1}, we have encountered the problem with under-estimated variances. The variances did improve when the sample size is very large and when the cure proportions are very small. In this manuscript, we propose to take each $\bm \theta^{(r)}$, $r=R^*+1,\cdots,R,$ and evaluate the observed data log-likelihood function. Then, we take that $\bm \theta^{(r)}$ as the estimate of $\bm \theta$ for which the log-likelihood function value is the maximum. 


\section{Simulation study}\label{sect5}

\subsection{Model fitting}\label{sect5.1}

{\hspace{7mm}}In order to assess the performance of the two estimation methods, we carry out a large simulation study.  For sample sizes $n$ = 200 and 400, we vary the lifetime distribution parameters, cure rates and censoring proportions. In addition, for simplicity, and as done in Balakrishnan and Pal (2016), we include a covariate effect $x$ in the form of $x = j$ for $j = 1,2,3,4$. From hereon in, we refer to the observations associated with covariate value $j$ as belonging to group $j$. We also link the cure rate $\pi_0$ to the covariate $x$ through the relation $\pi_0(x,\boldsymbol\beta)=\left\{1+e^{\beta_0+\beta_1x}\right\}^{-1}$. It is clear that the cure rate will differ from one group to another. In order to determine the values of the regression parameters, two cure rates need to be fixed. If we assume the cure rates to be monotone decreasing in the covariate, and fix two cure rates, we can then solve for the other two cure rates. With this purpose, we fix the values of $\pi_0(x=1,\boldsymbol\beta)$ (for group 1) and $\pi_0(x=4,\boldsymbol\beta)$ (for group 4) as $\pi_{01}$ and $\pi_{04}$, respectively. This results in the following expressions for the regression parameters $\beta_0$ and $\beta_1$ as:
\begin{equation} \label{eqbetas}
\begin{array}{rcl}
\beta_{1} &=& \frac{1}{3}\bigg[\log\bigg(\frac{1}{\pi_{04}}-1\bigg)-\log\bigg(\frac{1}{\pi_{01}}-1\bigg)\bigg] \\
  \beta_{0} &=& \log\bigg(\frac{1}{\pi_{01}}-1\bigg) - \beta_{1}.
  \end{array}
\end{equation}
Using \eqref{eqbetas}, the cure rates for groups 2 and 3 can be easily calculated as $\pi_{02}=\left\{1+e^{\beta_0+2\beta_1}\right\}^{-1}$ and $\pi_{03}=\left\{1+e^{\beta_0+3\beta_1}\right\}^{-1}$, respectively.\\

{\hspace{7mm}}For cure rates, we just consider two levels, which we refer to as ``High" and ``Low". Within our study, in the high setting, we fix groups 1 and 4's cure rates as 0.50 and 0.20, respectively, and in the low setting, we fix them as 0.40 and 0.10, respectively. Finally, as mentioned in Section \ref{sect3}, we allow for observations to be right censored. In order to incorporate this mechanism, we fix the overall censoring proportion for each group ($p_j, j=1,2,3,4$). In the high setting, these are fixed as (0.65,0.50,0.40,0.30) and in the low setting, (0.50,0.40,0.30,0.20). With these values, for each group, realized  censoring times can be generated by assuming they follow an exponential distribution with rate parameter $\gamma$, which, for fixed censoring proportion $p$ and cure rate $\pi_0$, can be found by solving the following equation:
\begin{eqnarray} \label{eqgamma}
\frac{p-\pi_0}{1-\pi_0} &=& P[Y> C |M > 0] \nonumber \\ 
 &=& \frac {P[Y>C, M=1]}{ P[M =1]} \nonumber \\
&=& \frac{1}{1-\pi_0}\int_{0}^{\infty}{S(x) \gamma e^{-\gamma x}dx}, 
\end{eqnarray}
where under our Bernoulli cure rate model, from (\ref{eq8.1}), $1-\pi_0=\frac{\nu}{1+\nu}$. Note that $S(\cdot)$ is the survival function of the EW distribution, as defined in (\ref{eq6}).\\

{\hspace{7mm}}From here, assuming the various cure rates and censoring rates have been predetermined for each group, the following steps are followed to generate the observed lifetime, $T$, under our model. First, a value of $M$ is generated from a Bernoulli distribution with $P[M=1] = 1-\pi_0$ and with it, a censoring time $C$ is generated from an exponential distribution with rate parameter $\gamma$. If $M$ = 0, this means there is no risk and the true lifetime is infinite with respect to the event of interest and so in this case, the observed lifetime is $T$ = $C$. If $M$ = 1, there is risk and so a true lifetime, $Y$, from the EW distribution is generated with parameters ($\alpha,k, \lambda$) and in this case, the observed lifetime is simply $T=\text{min}\left\{Y, C\right\}$. Finally, if $T = Y$, the right censoring indicator, $\delta$, is taken as 1, otherwise, it is taken as 0.\\

{\hspace{7mm}}For the parameters of the lifetime distribution $Y$, we consider three different parameter settings as  1: $(\alpha,k,\lambda)=(2,1,1.5)$, 2: (1,2,1.5) and 3: (1,1.5,0.5). For these choices of lifetime parameters, we consider different combinations of cure rates and sample sizes, resulting in the  12 settings as given in Table \ref{tab:cases}.\\

\begin{table}[h!] 
\caption{Parameter settings} \label{tab:cases}
\begin{center}
\begin{tabular}{|c|c|c|c|c|c|c|c|c|} \hline
Lifetime Parameter & $n$ & Cure Rate & $\beta_0$ & $\beta_1$& $\pi_{01}$&$\pi_{02}$&$\pi_{03}$&$\pi_{04}$ \\
\hline
Setting 1&200&Low&-0.192&0.597&0.400&0.268&0.168&0.100\\
&400&Low&-0.192&0.597&0.400&0.268&0.168&0.100\\
&200&High&-0.462&0.462&0.500&0.386&0.284&0.200\\
&400&High&-0.462&0.462&0.500&0.386&0.284&0.200\\
\hline
Setting 2&200&Low&-0.192&0.597&0.400&0.268&0.168&0.100\\
&400&Low&-0.192&0.597&0.400&0.268&0.168&0.100\\
&200&High&-0.462&0.462&0.500&0.386&0.284&0.200\\
&400&High&-0.462&0.462&0.500&0.386&0.284&0.200\\
\hline
Setting 3&200&Low&-0.192&0.597&0.400&0.268&0.168&0.100\\
&400&Low&-0.192&0.597&0.400&0.268&0.168&0.100\\
&200&High&-0.462&0.462&0.500&0.386&0.284&0.200\\
&400&High&-0.462&0.462&0.500&0.386&0.284&0.200\\
\hline
\end{tabular}
\end{center}
\end{table}

{\hspace{7mm}}For a given sample of observations, once the parameter estimates are obtained, with the goal to construct confidence intervals, we numerically approximate the hessian matrix. As to be seen in the tables, this will allow for the calculation of associated coverage probabilities. For each parameter setting, as considered in Table \ref{tab:cases}, we  generate $K$ = 500 samples using Monte Carlo simulation. Note that within the SEM algorithm, we choose $N=1500$ runs and use the first 500 as burn-in. For the EM algorithm, the tolerance $\epsilon$ is selected as 0.001. For both methods, for each of the five model parameters, as initial values, we randomly choose a value in the parameter space within 10\% of the true value.\\

{\hspace{7mm}}In Tables \ref{tab:params1}-\ref{tab:params3}, we summarize the performance of the two methods in estimating the model parameters. The tables include the estimates (and standard errors), bias, root mean square error (RMSE) and two coverage probabilities (90\% and 95\%). We first observe that as $n$ increases, with everything else fixed, the bias and RMSE both decrease, and coverage probabilities improve. Comparing the two methods, over all the settings, we observe that the two methods produce nearly the same results. Subsequently, in Tables \ref{tab:p0s1}-\ref{tab:p0s3}, we summarize the corresponding results for the estimation of cure rates. Across all parameter settings, it is evident that the estimates of the cure rates are consistently unbiased.

\setlength{\tabcolsep}{2pt}
%
\begin{table}[hptb]
\caption{Comparison of SEM and EM estimation results of model parameters when $(\alpha,\lambda,k)$ = (2,1.5,1) and the initial guess is close to the true parameter values} \label{tab:params1}
\begin{center}
\begin{tabular}{|c|l|c|c|c|c|c|c|c|c|c|c|} \hline
\multicolumn{1}{|c|}{$n$}&\multicolumn{1}{|c|}{Parameters}&\multicolumn{2}{|c|}{Estimates (SE)}&\multicolumn{2}{c|}{Bias}&\multicolumn{2}{c|}{RMSE}&\multicolumn{2}{c|}{90\% CP}&\multicolumn{2}{c|}{95\% CP}\\
\hline
&&\multicolumn{1}{|c|}{SEM}&\multicolumn{1}{|c|}{EM}&\multicolumn{1}{c|}{SEM}&\multicolumn{1}{c|}{EM}&\multicolumn{1}{c|}{SEM}&\multicolumn{1}{c|}{EM}&\multicolumn{1}{c|}{SEM}&\multicolumn{1}{c|}{EM}&\multicolumn{1}{c|}{SEM}&\multicolumn{1}{c|}{EM}\\ 
\hline
200&$\beta_0$=-0.192&-0.258(0.447)&-0.257(0.447)&-0.066&-0.065&0.480&0.479&0.874&0.874&0.936&0.936\\ 
&$\beta_1$=0.597&~0.641(0.204)&~0.640(0.204)&~0.043&~0.042&0.224&0.224&0.884&0.868&0.938&0.944\\ 
&$\alpha$=2&~2.138(1.383)&~2.139(1.226)&~0.138&~0.139&1.021&1.013&0.870&0.878&0.902&0.894\\ 
&$\lambda$=1.5&~1.590(0.710)&~1.586(0.642)&~0.090&~0.086&0.567&0.562&0.898&0.910&0.954&0.968\\ 
&$k$=1&~1.084(0.333)&~1.082(0.304)&~0.084&~0.082&0.295&0.293&0.948&0.966&0.982&0.998\\ 
\hline
400&$\beta_0$=-0.192&-0.205(0.313)&-0.206(0.313)&-0.013&-0.014&0.328&0.329&0.884&0.886&0.932&0.938\\ 
&$\beta_1$=0.597&~0.609(0.141)&~0.609(0.141)&~0.012&~0.012&0.144&0.145&0.898&0.908&0.950&0.950\\ 
&$\alpha$=2&~2.190(0.927)&~2.188(0.865)&~0.190&~0.188&0.881&0.881&0.868&0.870&0.910&0.904\\ 
&$\lambda$=1.5&~1.511(0.471)&~1.512(0.450)&~0.011&~0.012&0.466&0.466&0.884&0.880&0.930&0.932\\ 
&$k$=1&~1.028(0.207)&~1.028(0.199)&~0.028&~0.028&0.219&0.218&0.894&0.894&0.948&0.942\\ 
\hline
200&$\beta_0$=-0.462&0.453(0.438)&-0.449(0.438)&~0.009&~0.013&0.451&0.448&0.896&0.898&0.952&0.950\\ 
&$\beta_1$=0.462&~0.468(0.175)&~0.467(0.175)&~0.006&~0.005&0.175&0.174&0.908&0.912&0.952&0.952\\ 
&$\alpha$=2&~2.016(1.469)&~2.022(1.291)&~0.016&~0.022&1.056&1.053&0.808&0.804&0.842&0.840\\ 
&$\lambda$=1.5&~1.708(0.794)&~1.702(0.711)&~0.208&~0.202&0.686&0.682&0.908&0.910&0.946&0.958\\ 
&$k$=1&~1.154(0.402)&~1.151(0.364)&~0.154&~0.151&0.394&0.393&0.968&0.974&0.984&0.996\\ 
\hline
400&$\beta_0$=-0.462&-0.463(0.307)&-0.461(0.307)&-0.001&~0.001&0.320&0.320&0.892&0.888&0.942&0.938\\ 
&$\beta_1$=0.462&~0.462(0.123)&~0.461(0.123)&~0.000&-0.001&0.126&0.126&0.898&0.894&0.936&0.938\\ 
&$\alpha$=2&~2.144(1.079)&~2.139(0.940)&~0.144&~0.139&0.907&0.898&0.858&0.850&0.906&0.900\\ 
&$\lambda$=1.5&~1.547(0.545)&~1.549(0.503)&~0.047&~0.049&0.487&0.486&0.900&0.904&0.950&0.950\\ 
&$k$=1&~1.047(0.246)&~1.047(0.228)&~0.047&~0.047&0.232&0.230&0.932&0.932&0.974&0.974\\ 
\hline
\end{tabular}
\raggedright \footnotesize{\hspace{17mm} CP: coverage probability, SE: standard error}\\
\end{center}
\end{table}

\begin{table}[hptb]
\caption{Comparison of SEM and EM estimation results of model parameters when $(\alpha,\lambda,k)$ = (1,1.5,2) and the initial guess is close to the true parameter values} \label{tab:s2}
\begin{center}
\begin{tabular}{|c|l|c|c|c|c|c|c|c|c|c|c|} \hline
\multicolumn{1}{|c|}{$n$}&\multicolumn{1}{|c|}{Parameters}&\multicolumn{2}{|c|}{Estimates (SE)}&\multicolumn{2}{c|}{Bias}&\multicolumn{2}{c|}{RMSE}&\multicolumn{2}{c|}{90\% CP}&\multicolumn{2}{c|}{95\% CP}\\
\hline
&&\multicolumn{1}{|c|}{SEM}&\multicolumn{1}{|c|}{EM}&\multicolumn{1}{c|}{SEM}&\multicolumn{1}{c|}{EM}&\multicolumn{1}{c|}{SEM}&\multicolumn{1}{c|}{EM}&\multicolumn{1}{c|}{SEM}&\multicolumn{1}{c|}{EM}&\multicolumn{1}{c|}{SEM}&\multicolumn{1}{c|}{EM}\\ 
\hline
200&$\beta_0$=-0.192&-0.223(0.440)&-0.224(0.440)&-0.031&-0.032&0.452&0.451&0.902&0.906&0.946&0.948\\ 
&$\beta_1$=0.597&~0.622(0.199)&~0.622(0.199)&~0.024&~0.025&0.206&0.206&0.900&0.896&0.942&0.944\\ 
&$\alpha$=1&~1.097(0.525)&~1.097(0.503)&~0.097&~0.097&0.576&0.565&0.868&0.872&0.892&0.892\\ 
&$\lambda$=1.5&~1.501(0.292)&~1.500(0.281)&~0.001&~0.000&0.292&0.290&0.882&0.880&0.922&0.918\\ 
&$k$=2&~2.172(0.608)&~2.167(0.583)&~0.172&~0.167&0.692&0.687&0.884&0.894&0.948&0.944\\ 
\hline
400&$\beta_0$=-0.192&-0.231(0.308)&-0.230(0.308)&-0.039&-0.038&0.315&0.313&0.900&0.896&0.948&0.948\\ 
&$\beta_1$=0.597&~0.619(0.138)&~0.619(0.139)&~0.022&~0.022&0.147&0.146&0.902&0.900&0.948&0.950\\ 
&$\alpha$=1&~1.038(0.323)&~1.040(0.318)&~0.038&~0.040&0.346&0.347&0.878&0.874&0.916&0.914\\ 
&$\lambda$=1.5&~1.506(0.201)&~1.504(0.199)&~0.006&~0.004&0.208&0.208&0.884&0.886&0.946&0.940\\ 
&$k$=2&~2.080(0.389)&~2.077(0.383)&~0.080&~0.077&0.408&0.406&0.904&0.890&0.962&0.954\\ 
\hline
200&$\beta_0$=-0.462&-0.459(0.431)&-0.458(0.431)&~0.003&~0.004&0.456&0.456&0.886&0.888&0.956&0.954\\ 
&$\beta_1$=0.462&~0.467(0.173)&~0.467(0.173)&~0.005&~0.005&0.179&0.179&0.882&0.886&0.952&0.952\\ 
&$\alpha$=1&~1.143(0.653)&~1.141(0.590)&~0.143&~0.141&0.626&0.622&0.880&0.872&0.908&0.904\\ 
&$\lambda$=1.5&~1.482(0.338)&~1.483(0.311)&-0.018&-0.017&0.308&0.307&0.876&0.874&0.928&0.940\\ 
&$k$=2&~2.168(0.719)&~2.167(0.654)&~0.168&~0.167&0.728&0.725&0.904&0.900&0.950&0.954\\ 
\hline
400&$\beta_0$=-0.462&-0.500(0.303)&-0.500(0.303)&-0.037&-0.038&0.315&0.313&0.900&0.900&0.944&0.948\\ 
&$\beta_1$=0.462&~0.478(0.121)&~0.478(0.121)&~0.016&~0.016&0.128&0.128&0.880&0.886&0.926&0.924\\ 
&$\alpha$=1&~1.078(0.377)&~1.080(0.369)&~0.078&~0.080&0.393&0.392&0.910&0.906&0.924&0.926\\ 
&$\lambda$=1.5&~1.482(0.226)&~1.481(0.221)&-0.018&-0.019&0.224&0.223&0.902&0.894&0.932&0.932\\ 
&$k$=2&~2.060(0.437)&~2.056(0.426)&~0.060&~0.056&0.454&0.450&0.894&0.882&0.950&0.946\\ 
\hline
\end{tabular}
\raggedright \footnotesize{\hspace{17mm} CP: coverage probability, SE: standard error}\\
\end{center}
\end{table}

\begin{table}[hptb]
\caption{Comparison of SEM and EM estimation results of model parameters when $(\alpha,\lambda,k)$ = (1,0.5,1.5) and the initial guess is close to the true parameter values} \label{tab:params3}
\begin{center}
\begin{tabular}{|c|l|c|c|c|c|c|c|c|c|c|c|} \hline
\multicolumn{1}{|c|}{$n$}&\multicolumn{1}{|c|}{Parameters}&\multicolumn{2}{|c|}{Estimates (SE)}&\multicolumn{2}{c|}{Bias}&\multicolumn{2}{c|}{RMSE}&\multicolumn{2}{c|}{90\% CP}&\multicolumn{2}{c|}{95\% CP}\\
\hline
&&\multicolumn{1}{|c|}{SEM}&\multicolumn{1}{|c|}{EM}&\multicolumn{1}{c|}{SEM}&\multicolumn{1}{c|}{EM}&\multicolumn{1}{c|}{SEM}&\multicolumn{1}{c|}{EM}&\multicolumn{1}{c|}{SEM}&\multicolumn{1}{c|}{EM}&\multicolumn{1}{c|}{SEM}&\multicolumn{1}{c|}{EM}\\ 
\hline
200&$\beta_0$=-0.192&-0.208(0.443)&-0.205(0.443)&-0.016&-0.013&0.460&0.461&0.890&0.894&0.946&0.940\\ 
&$\beta_1$=0.597&~0.611(0.200)&~0.610(0.200)&~0.014&~0.013&0.212&0.211&0.894&0.898&0.942&0.942\\ 
&$\alpha$=1&~1.084(0.517)&~1.084(0.495)&~0.084&~0.084&0.460&0.456&0.886&0.892&0.892&0.922\\ 
&$\lambda$=0.5&~0.502(0.131)&~0.501(0.126)&~0.002&~0.001&0.119&0.118&0.926&0.930&0.922&0.958\\ 
&$k$=1.5&~1.598(0.459)&~1.594(0.442)&~0.098&~0.094&0.474&0.469&0.932&0.932&0.948&0.976\\ 
\hline
400&$\beta_0$=-0.192&-0.202(0.311)&-0.202(0.311)&-0.011&-0.010&0.307&0.306&0.904&0.904&0.948&0.960\\ 
&$\beta_1$=0.597&~0.605(0.140)&~0.605(0.140)&~0.008&~0.008&0.140&0.140&0.914&0.918&0.948&0.948\\ 
&$\alpha$=1&~1.056(0.339)&~1.056(0.334)&~0.056&~0.056&0.346&0.348&0.902&0.894&0.916&0.922\\ 
&$\lambda$=0.5&~0.501(0.092)&~0.501(0.090)&~0.001&~0.001&0.092&0.092&0.892&0.890&0.946&0.940\\ 
&$k$=1.5&~1.545(0.300)&~1.545(0.296)&~0.045&~0.045&0.316&0.317&0.896&0.890&0.962&0.944\\ 
\hline
200&$\beta_0$=-0.462&-0.488(0.436)&-0.488(0.436)&-0.026&-0.026&0.427&0.426&0.906&0.910&0.956&0.956\\ 
&$\beta_1$=0.462&~0.474(0.175)&~0.474(0.175)&~0.012&~0.012&0.178&0.178&0.904&0.902&0.952&0.952\\ 
&$\alpha$=1&~1.055(0.588)&~1.054(0.542)&~0.055&~0.054&0.451&0.447&0.896&0.888&0.908&0.924\\ 
&$\lambda$=0.5&~0.508(0.152)&~0.509(0.142)&~0.008&~0.009&0.122&0.122&0.928&0.938&0.928&0.974\\ 
&$k$=1.5&~1.639(0.558)&~1.637(0.523)&~0.139&~0.137&0.515&0.512&0.964&0.962&0.950&0.992\\ 
\hline
400&$\beta_0$=-0.462&-0.496(0.306)&-0.493(0.306)&-0.034&-0.031&0.313&0.313&0.894&0.890&0.944&0.942\\ 
&$\beta_1$=0.462&~0.477(0.123)&~0.476(0.123)&~0.015&~0.014&0.125&0.125&0.894&0.900&0.926&0.950\\ 
&$\alpha$=1&~1.061(0.391)&~1.061(0.375)&~0.061&~0.061&0.384&0.382&0.890&0.880&0.924&0.914\\ 
&$\lambda$=0.5&~0.500(0.102)&~0.500(0.099)&~0.000&~0.000&0.100&0.100&0.884&0.878&0.932&0.932\\ 
&$k$=1.5&~1.574(0.351)&~1.574(0.341)&~0.074&~0.074&0.380&0.379&0.898&0.892&0.950&0.952\\ 
\hline
\end{tabular}
\raggedright \footnotesize{\hspace{17mm} CP: coverage probability, SE: standard error}\\
\end{center}
\end{table}

\begin{table}[hptb]
\caption{Comparison of SEM and EM estimation results of cure rates when $(\alpha,\lambda,k)$ = (2,1.5,1) and the initial guess is close to the true parameter values} \label{tab:p0s1}
\begin{center}
\begin{tabular}{|c|l|c|c|c|c|c|c|c|c|c|c|} \hline
\multicolumn{1}{|c|}{$n$}&\multicolumn{1}{c|}{Parameters}&\multicolumn{2}{c|}{Estimates (SE)}&\multicolumn{2}{c|}{Bias}&\multicolumn{2}{c|}{RMSE}&\multicolumn{2}{c|}{90\% CP}&\multicolumn{2}{c|}{95\% CP}\\
\hline
&&\multicolumn{1}{c|}{SEM}&\multicolumn{1}{c|}{EM}&\multicolumn{1}{c|}{SEM}&\multicolumn{1}{c|}{EM}&\multicolumn{1}{c|}{SEM}&\multicolumn{1}{c|}{EM}&\multicolumn{1}{c|}{SEM}&\multicolumn{1}{c|}{EM}&\multicolumn{1}{c|}{SEM}&\multicolumn{1}{c|}{EM}\\ 
\hline
200&$\pi_{01}$=0.400&0.408(0.066)&0.407(0.066)&~0.008&~0.007&0.070&0.070&0.870&0.878&0.924&0.928\\ 
&$\pi_{02}$=0.268&0.266(0.039)&0.267(0.039)&-0.002&-0.002&0.040&0.040&0.902&0.892&0.944&0.946\\ 
&$\pi_{03}$=0.168&0.164(0.039)&0.164(0.038)&-0.004&-0.004&0.041&0.041&0.866&0.860&0.920&0.916\\ 
&$\pi_{04}$=0.100&0.099(0.038)&0.099(0.038)&-0.001&-0.001&0.041&0.041&0.818&0.822&0.886&0.890\\ 
\hline
400&$\pi_{01}$=0.400&0.401(0.047)&0.401(0.047)&~0.001&~0.001&0.050&0.050&0.870&0.880&0.940&0.938\\ 
&$\pi_{02}$=0.268&0.267(0.028)&0.267(0.028)&-0.001&-0.001&0.029&0.029&0.876&0.886&0.940&0.944\\ 
&$\pi_{03}$=0.168&0.167(0.028)&0.167(0.028)&-0.001&-0.001&0.028&0.028&0.878&0.892&0.948&0.950\\ 
&$\pi_{04}$=0.100&0.101(0.028)&0.100(0.028)&~0.001&~0.000&0.028&0.028&0.876&0.876&0.926&0.930\\ 
\hline
200&$\pi_{01}$=0.500&0.496(0.071)&0.496(0.071)&-0.004&-0.004&0.074&0.073&0.874&0.872&0.926&0.928\\ 
&$\pi_{02}$=0.386&0.383(0.045)&0.382(0.045)&-0.004&-0.004&0.047&0.047&0.874&0.876&0.916&0.922\\ 
&$\pi_{03}$=0.284&0.281(0.045)&0.281(0.045)&-0.003&-0.003&0.045&0.045&0.880&0.886&0.946&0.950\\ 
&$\pi_{04}$=0.200&0.201(0.055)&0.201(0.055)&~0.001&~0.001&0.056&0.055&0.882&0.888&0.944&0.940\\ 
\hline
400&$\pi_{01}$=0.500&0.500(0.050)&0.500(0.050)&~0.000&~0.000&0.052&0.052&0.876&0.880&0.926&0.928\\ 
&$\pi_{02}$=0.386&0.387(0.032)&0.387(0.032)&~0.001&~0.001&0.032&0.032&0.888&0.894&0.936&0.936\\ 
&$\pi_{03}$=0.284&0.285(0.032)&0.285(0.032)&~0.001&~0.001&0.032&0.032&0.908&0.902&0.948&0.948\\ 
&$\pi_{04}$=0.200&0.203(0.040)&0.203(0.040)&~0.003&~0.003&0.040&0.039&0.910&0.914&0.930&0.934\\ 
\hline
\end{tabular}
\raggedright \footnotesize{\hspace{17mm} CP: coverage probability, SE: standard error}\\
\end{center}
\end{table}

\begin{table}[hptb]
\caption{Comparison of SEM and EM estimation results of cure rates when $(\alpha,\lambda,k)$ = (1,1.5,2) and the initial guess is close to the true parameter values} \label{tab:p0s2}
\begin{center}
\begin{tabular}{|c|l|c|c|c|c|c|c|c|c|c|c|} \hline
\multicolumn{1}{|c}{$n$}&\multicolumn{1}{|c|}{Parameters}&\multicolumn{2}{|c|}{Estimates (SE)}&\multicolumn{2}{c|}{Bias}&\multicolumn{2}{c|}{RMSE}&\multicolumn{2}{c|}{90\% CP}&\multicolumn{2}{c|}{95\% CP}\\
\hline
&&\multicolumn{1}{|c|}{SEM}&\multicolumn{1}{|c|}{EM}&\multicolumn{1}{c|}{SEM}&\multicolumn{1}{c|}{EM}&\multicolumn{1}{c|}{SEM}&\multicolumn{1}{c|}{EM}&\multicolumn{1}{c|}{SEM}&\multicolumn{1}{c|}{EM}&\multicolumn{1}{c|}{SEM}&\multicolumn{1}{c|}{EM}\\
\hline
200&$\pi_{01}$=0.400&0.404(0.065)&0.404(0.065)&~0.004&~0.004&0.067&0.067&0.884&0.890&0.932&0.934\\ 
&$\pi_{02}$=0.268&0.267(0.038)&0.267(0.038)&-0.001&-0.001&0.040&0.040&0.892&0.894&0.934&0.934\\ 
&$\pi_{03}$=0.168&0.166(0.038)&0.166(0.038)&-0.002&-0.002&0.040&0.040&0.878&0.874&0.928&0.932\\ 
&$\pi_{04}$=0.100&0.102(0.038)&0.101(0.038)&~0.002&~0.001&0.041&0.041&0.876&0.872&0.912&0.910\\ 
\hline
400&$\pi_{01}$=0.400&0.405(0.046)&0.405(0.046)&~0.005&~0.005&0.047&0.046&0.886&0.882&0.946&0.948\\ 
&$\pi_{02}$=0.268&0.268(0.027)&0.268(0.027)&~0.000&~0.000&0.027&0.027&0.916&0.914&0.952&0.954\\ 
&$\pi_{03}$=0.168&0.166(0.027)&0.166(0.027)&-0.002&-0.002&0.028&0.028&0.882&0.880&0.928&0.930\\ 
&$\pi_{04}$=0.100&0.100(0.027)&0.099(0.027)&~0.000&-0.001&0.028&0.028&0.868&0.864&0.918&0.916\\ 
\hline
200&$\pi_{01}$=0.500&0.498(0.069)&0.498(0.069)&-0.002&-0.002&0.074&0.073&0.876&0.884&0.932&0.928\\ 
&$\pi_{02}$=0.386&0.385(0.044)&0.384(0.044)&-0.002&-0.002&0.045&0.045&0.884&0.878&0.946&0.948\\ 
&$\pi_{03}$=0.284&0.283(0.044)&0.283(0.044)&-0.001&-0.002&0.043&0.043&0.910&0.904&0.956&0.952\\ 
&$\pi_{04}$=0.200&0.202(0.054)&0.202(0.054)&~0.002&~0.002&0.054&0.054&0.890&0.894&0.940&0.938\\ 
\hline
400&$\pi_{01}$=0.500&0.505(0.049)&0.505(0.049)&~0.005&~0.005&0.050&0.050&0.900&0.906&0.938&0.940\\ 
&$\pi_{02}$=0.386&0.388(0.031)&0.388(0.031)&~0.002&~0.002&0.031&0.030&0.914&0.920&0.946&0.950\\ 
&$\pi_{03}$=0.284&0.283(0.031)&0.283(0.031)&-0.001&-0.001&0.032&0.032&0.890&0.892&0.936&0.942\\ 
&$\pi_{04}$=0.200&0.199(0.038)&0.199(0.038)&-0.001&-0.001&0.040&0.040&0.876&0.878&0.930&0.934\\ 
\hline
\end{tabular}
\raggedright \footnotesize{\hspace{17mm} CP: coverage probability, SE: standard error}\\
\end{center}
\end{table}

\begin{table}[hptb]
\caption{Comparison of SEM and EM estimation results of cure rates when $(\alpha,\lambda,k)$ = (1,0.5,1.5) and the initial guess is close to the true parameter values} \label{tab:p0s3}
\begin{center}
\begin{tabular}{|c|l|c|c|c|c|c|c|c|c|c|c|} \hline
\multicolumn{1}{|c|}{$n$}&\multicolumn{1}{c|}{Parameters}&\multicolumn{2}{c|}{Estimates (SE)}&\multicolumn{2}{c|}{Bias}&\multicolumn{2}{c|}{RMSE}&\multicolumn{2}{c|}{90\% CP}&\multicolumn{2}{c|}{95\% CP}\\
\hline
&&\multicolumn{1}{c|}{SEM}&\multicolumn{1}{c|}{EM}&\multicolumn{1}{c|}{SEM}&\multicolumn{1}{c|}{EM}&\multicolumn{1}{c|}{SEM}&\multicolumn{1}{c|}{EM}&\multicolumn{1}{c|}{SEM}&\multicolumn{1}{c|}{EM}&\multicolumn{1}{c|}{SEM}&\multicolumn{1}{c|}{EM}\\ 
\hline
200&$\pi_{01}$=0.400&0.403(0.066)&0.402(0.066)&~0.003&~0.002&0.067&0.067&0.880&0.880&0.940&0.938\\ 
&$\pi_{02}$=0.268&0.268(0.039)&0.268(0.039)&~0.000&-0.001&0.037&0.038&0.918&0.922&0.966&0.964\\ 
&$\pi_{03}$=0.168&0.168(0.038)&0.168(0.038)&~0.000&~0.000&0.039&0.038&0.896&0.902&0.946&0.948\\ 
&$\pi_{04}$=0.100&0.104(0.039)&0.104(0.039)&~0.004&~0.004&0.041&0.040&0.878&0.880&0.916&0.916\\ 
\hline
400&$\pi_{01}$=0.400&0.402(0.047)&0.401(0.047)&~0.002&~0.001&0.046&0.046&0.894&0.898&0.954&0.950\\ 
&$\pi_{02}$=0.268&0.268(0.027)&0.268(0.027)&~0.000&~0.000&0.028&0.027&0.892&0.894&0.944&0.948\\ 
&$\pi_{03}$=0.168&0.168(0.027)&0.168(0.027)&~0.000&~0.000&0.028&0.027&0.904&0.902&0.940&0.948\\ 
&$\pi_{04}$=0.100&0.102(0.028)&0.102(0.028)&~0.002&~0.002&0.027&0.027&0.904&0.896&0.936&0.934\\ 
\hline
200&$\pi_{01}$=0.500&0.504(0.070)&0.503(0.070)&~0.004&~0.003&0.068&0.068&0.906&0.914&0.950&0.948\\ 
&$\pi_{02}$=0.386&0.388(0.044)&0.388(0.044)&~0.002&~0.001&0.044&0.044&0.910&0.900&0.944&0.944\\ 
&$\pi_{03}$=0.284&0.285(0.045)&0.285(0.045)&~0.001&~0.000&0.047&0.047&0.892&0.892&0.940&0.940\\ 
&$\pi_{04}$=0.200&0.203(0.055)&0.203(0.055)&~0.003&~0.003&0.058&0.058&0.890&0.892&0.926&0.926\\ 
\hline
400&$\pi_{01}$=0.500&0.505(0.050)&0.504(0.050)&~0.005&~0.004&0.052&0.052&0.870&0.872&0.928&0.926\\ 
&$\pi_{02}$=0.386&0.388(0.031)&0.388(0.031)&~0.002&~0.001&0.034&0.034&0.870&0.866&0.932&0.928\\ 
&$\pi_{03}$=0.284&0.283(0.032)&0.283(0.032)&-0.001&-0.001&0.034&0.034&0.888&0.896&0.936&0.936\\ 
&$\pi_{04}$=0.200&0.199(0.039)&0.199(0.039)&-0.001&-0.001&0.041&0.040&0.882&0.892&0.942&0.940\\ 
\hline
\end{tabular}
\raggedright \footnotesize{\hspace{17mm} CP: coverage probability, SE: standard error}\\
\end{center}
\end{table}

\subsection{Robustness study with respect to the choice of initial values} \label{sect5.2}

{\hspace{7mm}}In this section, we study the robustness of the SEM and EM algorithms when the initial guess of the model parameters is far away from the true values. For this purpose, for each model parameter, we provide an initial guess that differs from its true value by at least 50\% and by at most 75\%. Then, we run the SEM and EM algorithms using the same choice of initial values to make sure that the comparison between the two algorithms is fair. In Table \ref{table:RI}, we present the percentage of divergent samples based on 500 Monte Carlo runs for different parameter settings. It is easy to see that for any considered parameter setting, the divergence percentage corresponding to the SEM algorithm is much less when compared to the EM algorithm. This clearly shows that the EM algorithm is sensitive to the choice of initial values, whereas the SEM algorithm is more robust. This, certainly, is a big advantage of the SEM algorithm and, hence, the SEM algorithm can be considered a preferred algorithm over the EM algorithm. It is interesting to note that when the true lifetime parameters are as considered in either setting 1 or setting 3, the percentage of divergent samples decrease with an increase in sample size. However, this is not true when the true lifetime parameters are as in setting 2. Similarly, for lifetime parameters as in settings 1 and 3, and for the SEM algorithm, the divergence percentages are smaller when the true cure rates are low. In this regard, for the EM algorithm, the divergence percentages are smaller for low cure rates, irrespective of the lifetime parameters.

\setlength{\tabcolsep}{6pt}.
\begin{table}[hptb]
\caption{Divergence rates (in \%) for the SEM and EM algorithms when the initial values deviate from the true values by at least 50\% and at most 75\%}\label{table:RI}
\begin{center}
\begin{tabular}{ccccc} \hline
$n$ & Cure Rate & Lifetime & \multicolumn{2}{c}{Divergence \%} \\ \cline{4-5}
 &  &  &  SEM & EM \\ \hline
 200 & High & Setting 1 & 6.800\% & 43.600\%\\
 400 & High & Setting 1 & 0.800\% & 37.600\%\\
 200 & Low & Setting 1 & 4.800\% & 39.000\%\\
 400 & Low & Setting 1 & 0.000\% & 38.200\%\\
\hline
 200 & High & Setting 2 & 5.255\% & 25.839\%\\
 400 & High & Setting 2 & 6.083\% & 25.816\%\\
 200 & Low & Setting 2 & 5.422\% & 24.397\%\\
 400 & Low & Setting 2 & 7.012\% & 23.780\%\\
\hline
 200 & High & Setting 3 & 4.282\% & 35.138\%\\
 400 & High & Setting 3  & 0.565\% & 29.378\%\\
 200 & Low & Setting 3  & 1.200\% & 32.200\%\\
 400 & Low & Setting 3  & 0.712\% & 28.632\%\\
\hline
\end{tabular}
\end{center}
\end{table}

\subsection{Robustness study with respect to the presence of outliers}\label{sect5.3}

{\hspace{7mm}}In this section, we study the performances of the SEM and EM algorithms when there are outliers present in the data. We consider a scenario where the generated data contains 5\% outliers. For this purpose, we generate 95\% of the data with true parameter setting as follows: $(\beta_0,\beta_1,\alpha,\lambda,k)$ = (-0.462,0.462,1,1.5,2), which corresponds to $(\pi_{01},\pi_{04})$ = (0.5,0.2) and $(p1,p2,p3,p4)$ = (0.65,0.50,0.40,0.30). The remaining 5\% of the data are outliers and are generated from $(\beta_0,\beta_1,\alpha,\lambda,k)$ = (-0.192,0.597,1,1,0.3), which corresponds to $(\pi_{01},\pi_{04})$ = (0.4,0.1) and $(p1,p2,p3,p4)$ = (0.50,0.40,0.30,0.20). As far as the lifetime distribution is concerned, the true parameter setting results in a mean of 1.329, whereas the setting to generate outliers result in a mean of 9.260. Then, for the entire data, we use both SEM and EM algorithms to estimate the true models parameters $(\beta_0,\beta_1,\alpha,\lambda,k)$ = (-0.462,0.462,1,1.5,2) and the true cure rates $(\pi_{01},\pi_{02},\pi_{03},\pi_{04})$ = (0.5,0.386,0.284,0.2). Based on 500 Monte Carlo simulations, we present the estimation results of model parameters and cure rates in Table \ref{table:O1} and Table \ref{table:O2}, respectively. From Table \ref{table:O1}, it is clear that the presence of outliers result in biased estimates, which is more pronounced for the lifetime parameters and specifically for the parameter $k$. This is certainly due to the choice of the parameters using which the outliers were generated. Note that there is also a significant under-coverage that can be noticed for the lifetime parameters. The increase in sample size helps in the reduction of the standard errors and the RMSEs. It also helps in the reduction of bias for all model parameters except for the parameter $k$. From Table \ref{table:O2}, we note that the estimates of cure rates contain little bias when compared to the results in Section \ref{sect5.1} where there was no outliers. A slight under-coverage is also noticed for some cure rates. Finally, the performances of the SEM and EM algorithms are similar in the presence of outliers.

\setlength{\tabcolsep}{3pt}
\begin{table}[hptb]
\caption{Comparison of SEM and EM estimation results of model parameters in the presence of outliers in the data}\label{table:O1}
\begin{center}
\begin{tabular}{|c|l|c|c|c|c|c|c|c|c|c|c|} \hline
\multicolumn{1}{|c|}{$n$}&\multicolumn{1}{|c|}{Parameters}&\multicolumn{2}{c|}{Estimates (SE)}&\multicolumn{2}{c|}{Bias}&\multicolumn{2}{c|}{RMSE}&\multicolumn{2}{c|}{90\% CP}&\multicolumn{2}{c|}{95\% CP}\\
\hline
&&\multicolumn{1}{c|}{SEM}&\multicolumn{1}{c|}{EM}&\multicolumn{1}{c|}{SEM}&\multicolumn{1}{c|}{EM}&\multicolumn{1}{c|}{SEM}&\multicolumn{1}{c|}{EM}&\multicolumn{1}{c|}{SEM}&\multicolumn{1}{c|}{EM}&\multicolumn{1}{c|}{SEM}&\multicolumn{1}{c|}{EM}\\ 
\hline
200&$\beta_0$=-0.462&-0.422(0.452)&-0.422(0.452)&~0.040&~0.040&0.464&0.462&0.904&0.906&0.952&0.952\\ 
&$\beta_1$=0.462&~0.473(0.172)&~0.473(0.172)&~0.011&~0.011&0.177&0.176&0.894&0.898&0.942&0.948\\ 
&$\alpha$=1&~1.119(0.343)&~1.121(0.340)&~0.119&~0.121&0.776&0.781&0.528&0.520&0.622&0.616\\ 
&$\lambda$=1.5&~1.573(0.272)&~1.572(0.269)&~0.073&~0.072&0.453&0.454&0.610&0.604&0.722&0.708\\ 
&$k$=2&~1.750(0.367)&~1.748(0.359)&-0.250&-0.252&0.845&0.844&0.456&0.448&0.506&0.510\\ 
\hline
400&$\beta_0$=-0.462&-0.443(0.316)&-0.443(0.316)&~0.020&~0.019&0.311&0.309&0.894&0.900&0.952&0.954\\ 
&$\beta_1$=0.462&~0.473(0.120)&~0.473(0.120)&~0.011&~0.011&0.119&0.118&0.912&0.910&0.948&0.954\\ 
&$\alpha$=1&~1.105(0.213)&~1.106(0.212)&~0.105&~0.106&0.592&0.595&0.430&0.424&0.504&0.502\\ 
&$\lambda$=1.5&~1.528(0.181)&~1.528(0.181)&~0.028&~0.028&0.370&0.371&0.530&0.532&0.608&0.602\\ 
&$k$=2&~1.576(0.193)&~1.576(0.194)&-0.424&-0.424&0.748&0.750&0.256&0.264&0.326&0.324\\ 
\hline
\end{tabular}
\raggedright \footnotesize{\hspace{17mm} CP: coverage probability, SE: standard error}\\
\end{center}
\end{table}

\begin{table}[hptb]
\caption{Comparison of SEM and EM estimation results of cure rates in the presence of outliers in the data} \label{table:O2}
\begin{center}
\begin{tabular}{|c|l|c|c|c|c|c|c|c|c|c|c|} \hline
\multicolumn{1}{|c|}{$n$}&\multicolumn{1}{|c|}{Cure Rates}&\multicolumn{2}{|c|}{Estimates (SE)}&\multicolumn{2}{c|}{Bias}&\multicolumn{2}{c|}{RMSE}&\multicolumn{2}{c|}{90\% CP}&\multicolumn{2}{c|}{95\% CP}\\
\hline
&&\multicolumn{1}{|c|}{SEM}&\multicolumn{1}{|c|}{EM}&\multicolumn{1}{c|}{SEM}&\multicolumn{1}{c|}{EM}&\multicolumn{1}{c|}{SEM}&\multicolumn{1}{c|}{EM}&\multicolumn{1}{c|}{SEM}&\multicolumn{1}{c|}{EM}&\multicolumn{1}{c|}{SEM}&\multicolumn{1}{c|}{EM}\\
\hline
200&$\pi_{01}$=0.500&0.488(0.074)&0.488(0.074)&-0.012&-0.012&0.077&0.077&0.866&0.888&0.938&0.940\\ 
&$\pi_{03}$=0.386&0.373(0.046)&0.373(0.046)&-0.013&-0.013&0.050&0.050&0.796&0.878&0.926&0.922\\ 
&$\pi_{03}$=0.284&0.272(0.042)&0.272(0.042)&-0.012&-0.012&0.046&0.046&0.818&0.864&0.916&0.924\\ 
&$\pi_{04}$=0.200&0.192(0.050)&0.193(0.050)&-0.008&-0.007&0.054&0.054&0.828&0.868&0.920&0.922\\ 
\hline
400&$\pi_{01}$=0.500&0.493(0.052)&0.493(0.052)&-0.007&-0.007&0.053&0.052&0.844&0.890&0.954&0.952\\ 
&$\pi_{02}$=0.386&0.377(0.033)&0.377(0.033)&-0.009&-0.009&0.035&0.035&0.682&0.880&0.938&0.934\\ 
&$\pi_{03}$=0.284&0.275(0.030)&0.275(0.030)&-0.009&-0.009&0.032&0.032&0.724&0.854&0.932&0.932\\ 
&$\pi_{04}$=0.200&0.193(0.036)&0.193(0.036)&-0.007&-0.007&0.037&0.037&0.788&0.874&0.934&0.932\\ 
\hline
\end{tabular}
\raggedright \footnotesize{\hspace{17mm} CP: coverage probability, SE: standard error}\\
\end{center}
\end{table}

\subsection{Model discrimination}\label{sect5.4}
{\hspace{7mm}} As mentioned in Sections \ref{sect1} and \ref{sect2}, the EW family of distributions includes many well-known lifetime distributions. Consequently, it makes sense to carry out a model discrimination study across the sub-models through the general EW distribution. The idea is to evaluate the performance of the likelihood ratio test in discriminating among the sub models. For this purpose, we choose the setting with ``Low" cure rates and a sample of size 400. The EW scale parameter $\lambda$ is chosen to be $2.5$. \\

{\hspace{7mm}} Data from the Bernoulli cure rate model are generated with the lifetimes coming from the five special cases (true models) of the EW family, namely, exponential ($\alpha=1, k=1$), Rayleigh ($\alpha=1, k=2$), Weibull ($\alpha=1, k=1.5$), generalized exponential ($\alpha=2, k=1$) and Burr Type X ($\alpha=2, k=2$) distributions. For data generated from every true model, all five sub models are fitted and parameter estimation is carried out by applying the SEM algorithm specified in Section \ref{sect4.2}. In particular, we carry out following hypothesis tests corresponding to the five sub-models:
\begin{enumerate}
\item [-] Exponential: $H_{0}: \alpha=k=1$ vs. $H_{1}$: at least one inequality in $H_{0}$;
 \item [-] Rayleigh: $H_{0}: \alpha=1, k=2$ vs. $H_{1}$: at least one inequality in $H_{0}$;
\item [-] Weibull: $H_{0}: \alpha=1$ vs. $H_{1}: \alpha \ne 1$;
\item [-] Generalized exponential: $H_{0}: k=1$ vs. $H_{1}: k \ne 1$;
\item [-] Burr Type X: $H_{0}: k=2$ vs. $H_{1}: k \ne 2$.
\end{enumerate} 

Let $\hat l$ and $\hat l_0$ denote the unrestricted maximized log-likelihood value and the maximized log-likelihood value obtained under $H_0$, respectively. Then, by Wilk's theorem, $\Lambda=-2\left(\hat l - \hat l_0\right) {\sim} \chi^2_{q^*}$ asymptotically under $H_0$ where $\chi^2_{q^*}$ represents a chi-squared distribution with $q^*$ degrees of freedom and $q^*$ denotes the difference in the number of parameters estimated to obtain $\hat l$ and $\hat l_0$. The p-values for the tests are compared against a significance level of $0.05$ to decide whether to reject $H_0$ or not. Proportion of rejections of $H_0$ based on 1000 Monte Carlo runs are reported in Table \ref{MD} for every combination of the true and fitted models.   \\

\setlength{\tabcolsep}{3pt}
\begin{table}[h!]
    \centering
   \caption{Rejection rates of the true models based on likelihood ratio test}   
   \begin{tabular}{cccccc}
\cline{2-6}
&\multicolumn{5}{ c }{\bf True Model ($\lambda=2.5$)}\\    
\hline
{\bf Fitted Model} & Exponential & Rayleigh & Weibull & Generalized Exponential & Burr Type X\\ 
& & &($k=1.5$) & ($\alpha=2$)&($\alpha=2$)\\
\hline
Exponential &{\bf 0.044}&1.000&1.000&1.000&1.000\\
Rayleigh &1.000&{\bf 0.050}&1.000&1.000&1.000\\
Weibull & 0.052&0.040&{\bf 0.068}&0.542&0.596\\
Generalized Exponential & 0.054&0.964&0.578&{\bf 0.048}&0.950\\
Burr Type X & 0.904&0.042&0.330&0.956&{\bf 0.046}\\
\hline
    \end{tabular}
    \label{MD}
\end{table}

{\hspace{7mm}} The observed significance level corresponding to every true lifetime distribution is close to the nominal level or significance level of 0.05. This implies that the chi-squared distribution provides a good approximation to the null distribution of the likelihood ratio test statistic. Next, we observe that when the true lifetime model is either exponential or Rayleigh, the rejection rates for the fitted Weibull model are 0.052 and 0.040, respectively. These rejection rates are close to 0.05 because both exponential and Rayleigh are contained within the Weibull distribution. On the other hand, when the true lifetime is generalized exponential or Burr Type X, the rejection rates for the fitted Weibull lifetime are 0.542 and 0.596, respectively. These rejection rates are moderate because the Weibull distribution doesn't accommodate the generalized exponential or Burr Type X distributions as special cases. Based on some of the high rejection rates, we can conclude that the likelihood ratio test can discriminate between the following models: exponential and Rayleigh, Burr Type X and exponential, Burr Type X and generalized exponential, and generalized exponential and Rayleigh. As such, for a given data, there is a necessity to employ the likelihood ratio test for choosing the correct sub model, if possible. If none of the sub models provide an adequate fit, the proposed EW model should be used.

\section{Analysis of cutaneous melanoma data}\label{sect6}

\begin{figure}[h!]
\centering
\includegraphics[scale=0.3]{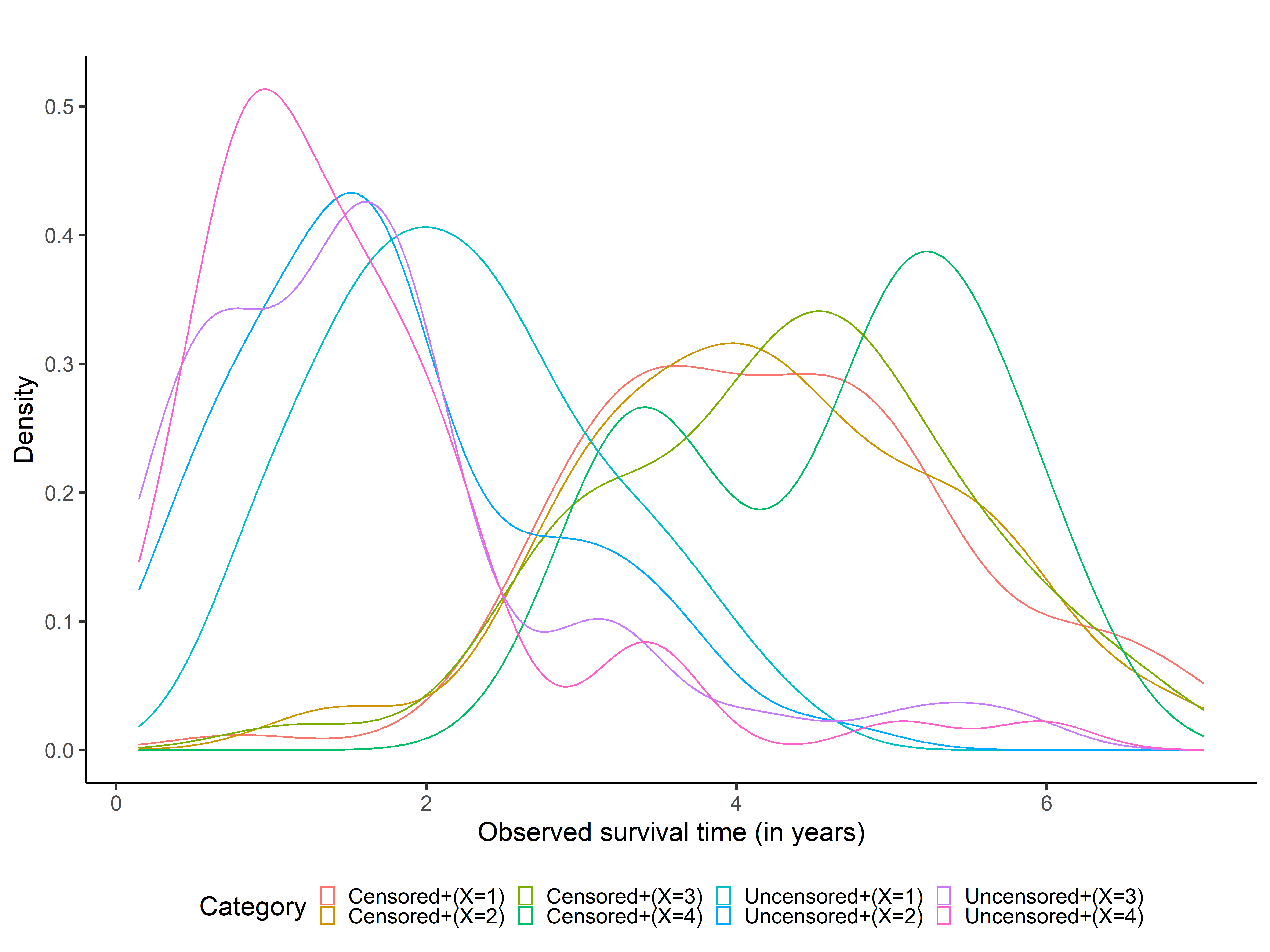}
\caption{Kernel density plots of the observed survival times categorized by censoring indicator and nodule category }
\label{Fig1}
\end{figure}

{\hspace{7mm}} {\bf {\it Data description}}: An illustration of our proposed model with EW lifetime and proposed estimation technique is presented in this section. Motivated by an example provided in \cite{ibrahim2005bayesian} which showed influences of cure fraction, we consider the data set on cutaneous melanoma (a type of malignant skin cancer) studied by the Eastern Cooperative Oncology Group (ECOG) where the patients were observed for the period between 1991-1995. The objective of the study was to assess the efficacy of the postoperative treatment with high dose of interferon alpha-2b drug to prevent recurrences of the cancer.  Observed survival time ($t$, in years) representing either exact lifetime or censoring time, censoring indicator ($\delta =0, 1$) and nodule category ($x=1, 2, 3, 4$) based on tumor thickness are selected as the variables of interest for demonstrating the performance of our model.  There are 427 observations in the data set; each observation corresponds to a patient in the study with respective nodule category information. Analysis is performed based on 417 patients' data due to missing information on tumor thickness for the remaining 10 patients. Nodule category is taken as the only covariate for our illustration. A descriptive summary of the observed survival time is given in Table \ref{labint} and kernel density plots of the same are given in Figure \ref{Fig1} categorized by censoring indicator and nodule category. \\

\begin{table}[hptb!]
    \centering
   \caption{Descriptive summary of the observed survival times ($t$) categorized by censoring indicator ($\delta$) and nodule category ($x$)}   
   \begin{tabular}{clcccc}
\hline
{\bf $\delta$} & Measure & $x=1$ & $x=2$ & $x=3$ & $x=4$\\ 
\hline
&$n$ (\%) & 36 (8.633) & 53 (12.709) & 41 (9.832) & 55 (13.189)\\
Uncensored&Mean& 2.227 & 1.777 & 1.687 & 1.551\\
($\delta=1$)&SD &0.889&1.006 &1.248&1.114\\
185 ($44.365\%$) &Median&2.185&1.599&1.544&1.268\\
 &Min&0.767&0.285&0.148&0.170\\
&Max&4.263&4.613&5.739&5.969\\
\hline
&$n$ (\%) & 75 (17.985) & 84  (20.143)& 46  (11.031) & 27  (6.474)\\
Censored &Mean& 4.265 & 4.217 & 4.312 & 4.609 \\
($\delta=0$)  &SD& 1.209 & 1.192 & 1.167 & 1.008\\
232 ($55.635\%$) &Median& 4.208 & 4.112 & 4.474 & 5.043\\
 &Min& 0.791 & 1.325 & 1.139 & 2.927\\
&Max& 7.012 & 6.976 & 6.623 & 6.045\\
\hline
&$n$ (\%) & 111 (26.618) & 137  (32.853)& 87  (20.863) & 82  (19.664)\\
Combined &Mean& 3.604 & 3.273 & 3.075 & 2.558 \\
417 ($100\%$)  &SD& 1.467 & 1.636 & 1.782 & 1.801\\
  &Median& 3.537 & 3.387 & 3.047 & 1.966\\
 &Min& 0.767 & 0.285 & 0.148 & 0.170\\
&Max& 7.012 & 6.976 & 6.623 & 6.045\\
\hline
    \end{tabular}
    \label{labint}\\
\raggedright \footnotesize{\hspace{17mm} $n$: sample size, SD: standard deviation, Min: Minimum, Max: Maximum }\\
\end{table}

\begin{figure}[h!]
\centering
\includegraphics[scale=0.6]{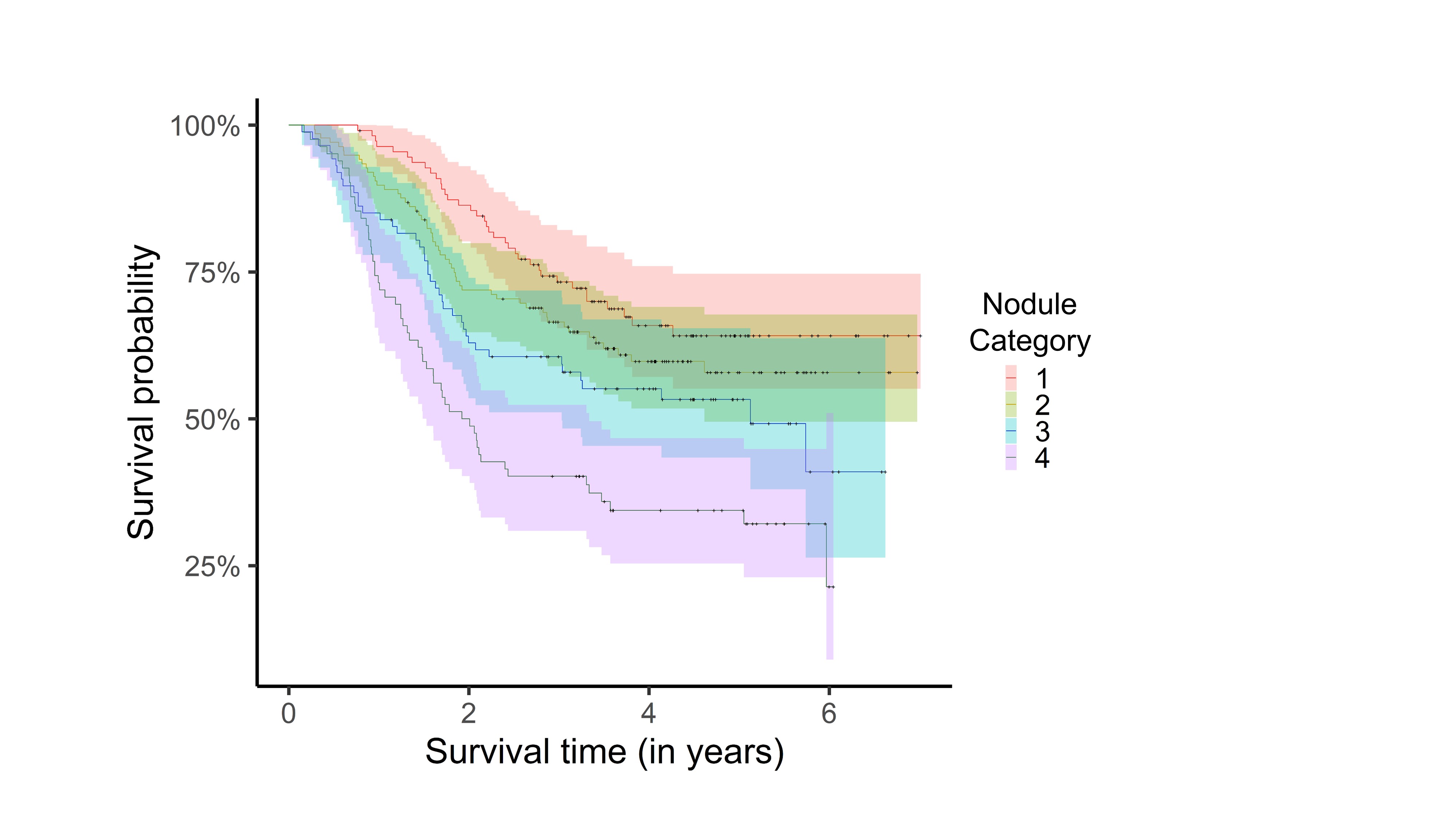}
\caption{Kaplan-Meier plots categorized by nodule category }
\label{Fig0}
\end{figure}

{\hspace{7mm}} {\bf {\it Assignment of initial parameter values }}: Let us define $\pi_{0x}, x=1, 2, 3, 4$, to be the cure rate for the $x$th nodule category. As indicated before and as can also be seen from Table \ref{labint}, $x=1$ represents the group which is likely to have the best prognosis (i.e., highest cure rate), whereas $x=4$ represents the group likely to have the worst prognosis (i.e., lowest cure rate). Further, the censoring rates for $x=1$ and $x=4$ are $0.675$ and $0.329$, respectively.  Using the monotone nature of the logistic-link function and assuming all censored individuals are cured, we obtain initial estimates of $\beta_0$ and $\beta_1$ by simultaneously solving
\begin{flalign*}\label{eq6.1}
&\pi_{01}=\left[1+\exp\{\beta_0+\beta_1\}\right]^{-1}=0.675\\ 
&\pi_{04}=\left[1+\exp\{\beta_0+4\beta_1\}\right]^{-1}=0.329. 
\end{flalign*}  
Hence, the initial estimates $\beta_0^{(0)}$ and $\beta_1^{(0)}$ are $-1.212$ and $0.481$, respectively. On the other hand, the initial estimates for the EW lifetime parameters $\alpha, \lambda$ and $k$ are obtained in two-steps. In the first step, we consider (\ref{eq6}) and note that
\begin{equation}\label{eq6.2}
\psi(t_i; \alpha, \lambda, k)=\log\left[-\log\left\{1-\left(1-S_s(t_i; \alpha, \lambda, k)\right)^{1/\alpha}\right\}\right] = k \log t_i - k \log \lambda
\end{equation} 
is linear in $t_i$, where $i \in \Delta_1$. Hence, fixing $\alpha=\alpha_0$, ordinary least square estimates  $\lambda_0$ of $\lambda$ and $k_0$ of $k$ are obtained by fitting a simple linear regression model with $\hat \psi(t; \alpha, \lambda, k)$  as the response and $t$ as the predictor. Here, $\hat \psi(t_i; \alpha, \lambda, k)=\log\left[-\log\left\{1-\left(1-\hat S_s(t_i; \alpha, \lambda, k)\right)^{1/\alpha}\right\}\right]$ and $\hat S_s(t_i; \alpha, \lambda, k)$ is the Kaplan-Meier estimate of the survival function evaluated at $t_i$ for the $i$th individual with $i \in \Delta_1$. In our case, $\alpha_0$ is chosen as 2. The Kaplan-Meier plots of the survival probabilities for the four nodule categories are presented in Figure \ref{Fig0}. In the second step, using (\ref{eq4}), we define a likelihood function $L_s(\alpha, \lambda, k; \bm t^*)$ as
\begin{equation}\label{eq6.3}
L_s(\alpha, \lambda, k; \bm t^*)=\prod_{i \in \Delta_1} f_s(t_i; \alpha, k, \lambda)= \prod_{i \in \Delta_1}\left(\frac{\alpha k}{\lambda}\right) \left(\frac{t_i}{\lambda}\right)^{k-1} e^{-(t_i/\lambda)^k}\left[1- e^{-(t_i/\lambda)^k}\right]^{\alpha-1},
\end{equation} 
where $\bm t^*=\{(t_i; i \in \Delta_1)\}$. From here, $\log L_s(\alpha, \lambda, k; \bm t^*)$ is then maximized with respect to $\alpha, \lambda$ and $k$ using numerical optimization routine in R with $\alpha_0$, $\lambda_0$ and $k_0$ as initial parameter guesses.  Finally, the ML estimates of $\alpha$, $\lambda$ and $k$ are obtained as  $\alpha^{(0)}=1.983,  \lambda^{(0)}=1.326$ and $ k^{(0)}=1.214$. Hence, $\bm \theta^{(0)}=\left(\beta_0^{(0)}, \beta_1^{(0)}, \alpha^{(0)}, \lambda^{(0)}, k^{(0)}\right) = (-1.212, 0.481, 1.983, 1.326, 1.214)$ is taken as the initial parameter guess for starting the iterative processes involved in both EM and SEM algorithms.\\

\begin{figure}[h!]
\centering
\includegraphics[scale=0.6]{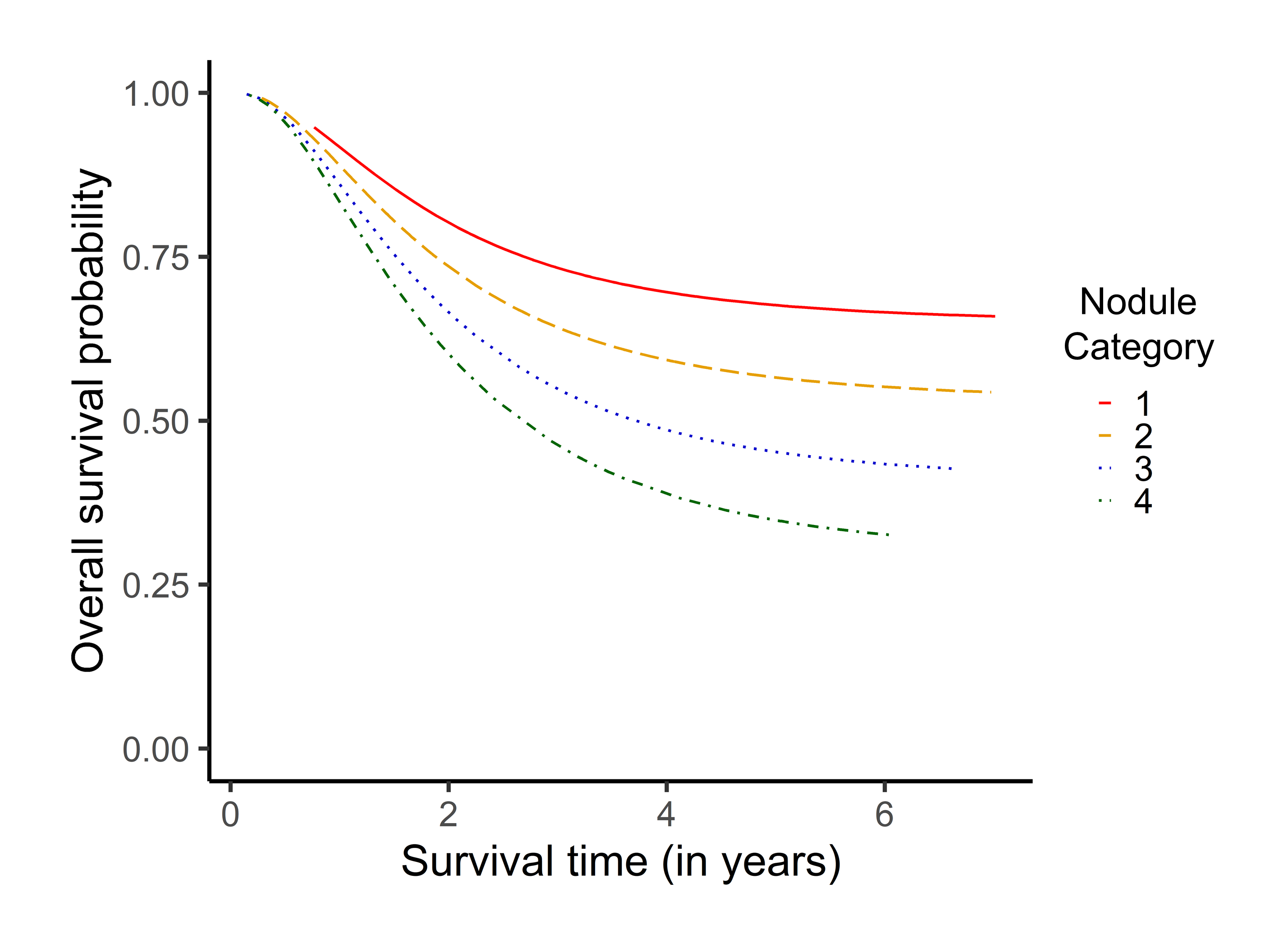}
\caption{Overall population survival probability plot  estimated point-wise by the SEM technique and categorized by nodule category }
\label{Fig3}
\end{figure}

{\hspace{7mm}} {\bf {\it Model fitting}}: As discussed in Section \ref{sect4}, model parameters are estimated by both EM and SEM methods. Point estimate, standard error (SE) and 95\% confidence interval (CI) are displayed in Table \ref{labint2} for both model parameters and cure rates for all nodules categories. The results clearly suggest that both methods of estimation provide similar values though standard errors are larger when estimation is done by  the SEM technique. This is quite expected due to the involvement of random number generation in SEM algorithm which adds to the variability of parameter estimates. The standard errors of the cure rate  estimates are estimated using the delta method. No overlap is observed between the confidence intervals for $\pi_{01}$ and $\pi_{04}$ suggesting that cure rates for these groups are significantly different.  Figure \ref{Fig3} presents plots corresponding to the overall population survivor function $S_p(.; \bm \theta)$, where $$\hat S_p(t_i; \bm \theta)=\left[\frac{1+e^{\hat \beta_0+ \hat \beta_1 x_i}\left\{1-\left[1-e^{-(t_i/\hat \lambda)^{\hat k}}\right]^{\hat \alpha}\right\}}{1+e^{\hat \beta_0+ \hat \beta_1 x_i}}\right]$$
evaluated at  observed $t_i$ for $i =1, \dots, n$. The plot shows similar pattern as that of the Kaplan-Meier plot in Figure \ref{Fig0}. It can be seen that overall survival probability plots level off to points much higher than 0 (even when patients were followed up for more than 6 years), therefore, strongly indicating the presence of significant cure fractions.\\

\begin{table}[hptb!]
    \centering
   \caption{Estimates, standard errors and 95\% confidence intervals of the model parameters and cure rates by applying EM and SEM algorithm on the cutaneous melanoma data set}   
   \begin{tabular}{cccc|ccc}
\hline
&\multicolumn{3}{ c |}{EM Algorithm ($\hat l = -513.836$)}&\multicolumn{3}{ c }{SEM Algorithm ($\hat l = -513.839$)}\\    
\hline
Parameter &	Estimate	&SE	&95\% CI	&  Estimate & 	SE &	95\% CI\\
\hline
$\beta_0$	& -1.114	&0.281	& (-1.666, -0.564)		& -1.117	& 0.283	& (-1.672, -0.562)\\
$\beta_1$	& 0.489	& 0.111	& (0.269, 0.709)	 	& 0.495	&0.113	&(0.274, 0.716)\\
$\alpha$	& 4.777	& 3.943	& (0.000, 12.505)	& 5.009	&4.704	&(0.000, 14.232)\\
$\lambda$	& 0.656	& 0.596	& (0.000, 1.825)	 	& 0.623	&0.657	&(0.000, 1.914)\\
$k$	&0.705	& 0.280	& (0.156, 1.254)       &0.688	& 0.308	& (0.083, 1.295)\\
\hline
Cure rate	 &Estimate	 & SE	&95\% CI		&Estimate	&SE	&95\% CI\\
\hline
$\pi_{01}$	&0.652	&0.045	&(0.564, 0.740)		&0.6507	&0.045	&(0.563, 0.739)\\
$\pi_{02}$	&0.534	&0.039	&(0.458, 0.610)		&0.5318	&0.041	&(0.451, 0.612)\\
$\pi_{03}$	&0.412	&0.045	&(0.324, 0.500)	        &0.4091	&0.048	&(0.315, 0.503)\\
$\pi_{04}$	&0.301	&0.055	&(0.193, 0.409)		&0.2967	&0.058	&(0.184, 0.410)\\
\hline
    \end{tabular}
    \label{labint2}\\
\raggedright \footnotesize{\hspace{17mm}  $\hat l$: Maximized log-likelihood value, SE: standard error, CI: confidence interval}\\
\end{table}

\begin{figure}[hptb!]
\centering
  \begin{tabular}{@{}cc@{}}
    \includegraphics[width=.45\textwidth]{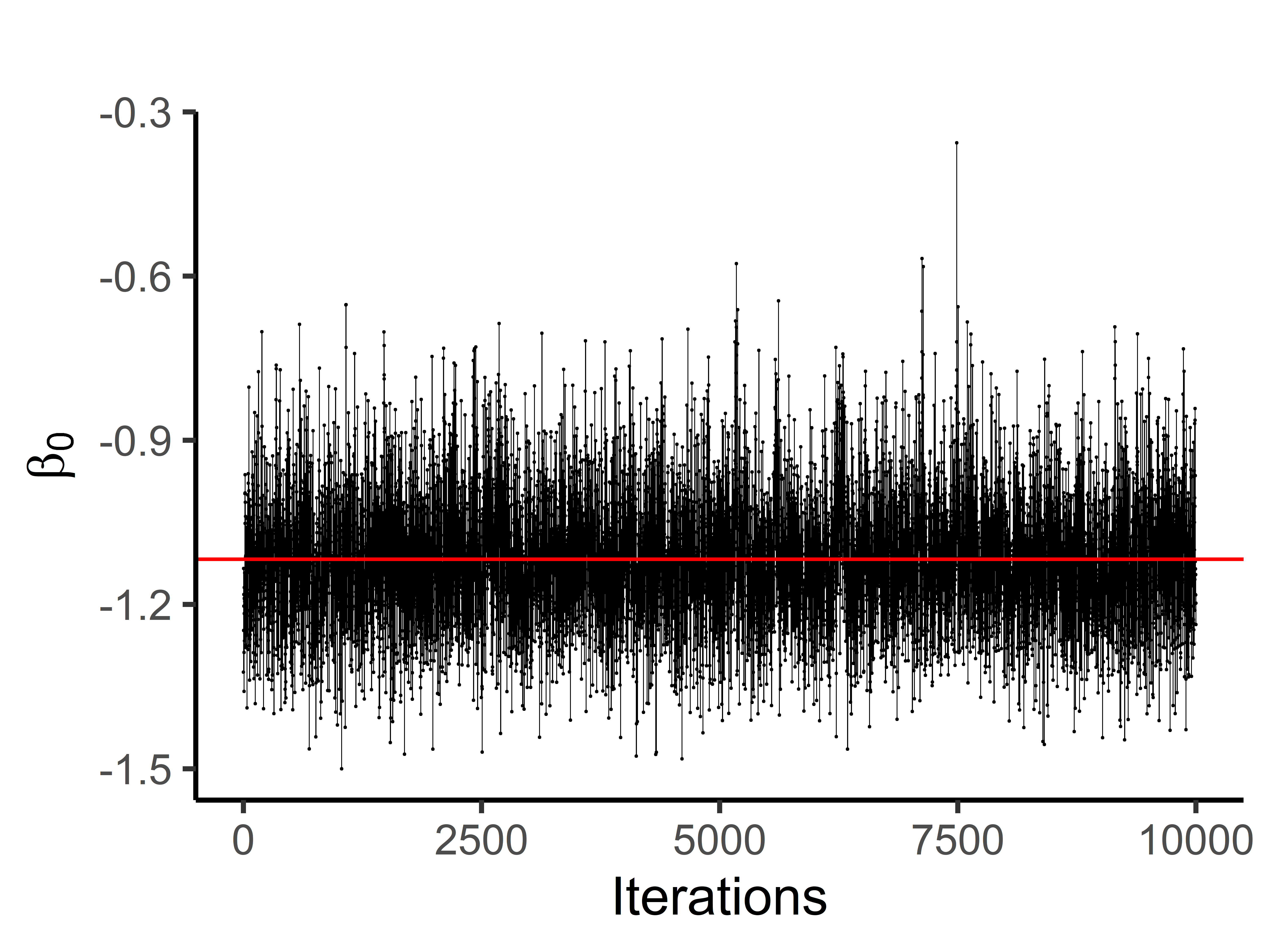} &
    \includegraphics[width=.45\textwidth]{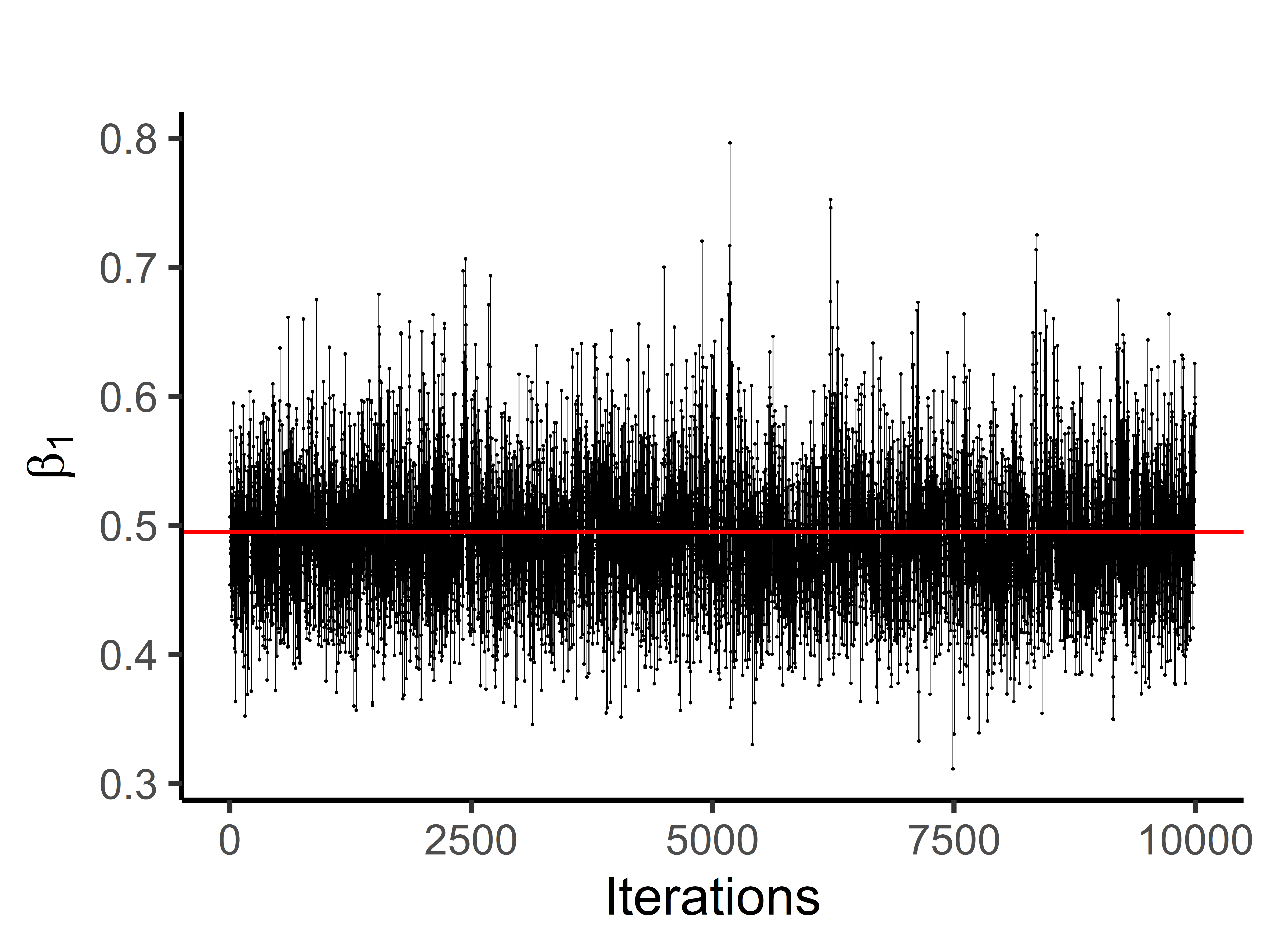} \\
    \includegraphics[width=.45\textwidth]{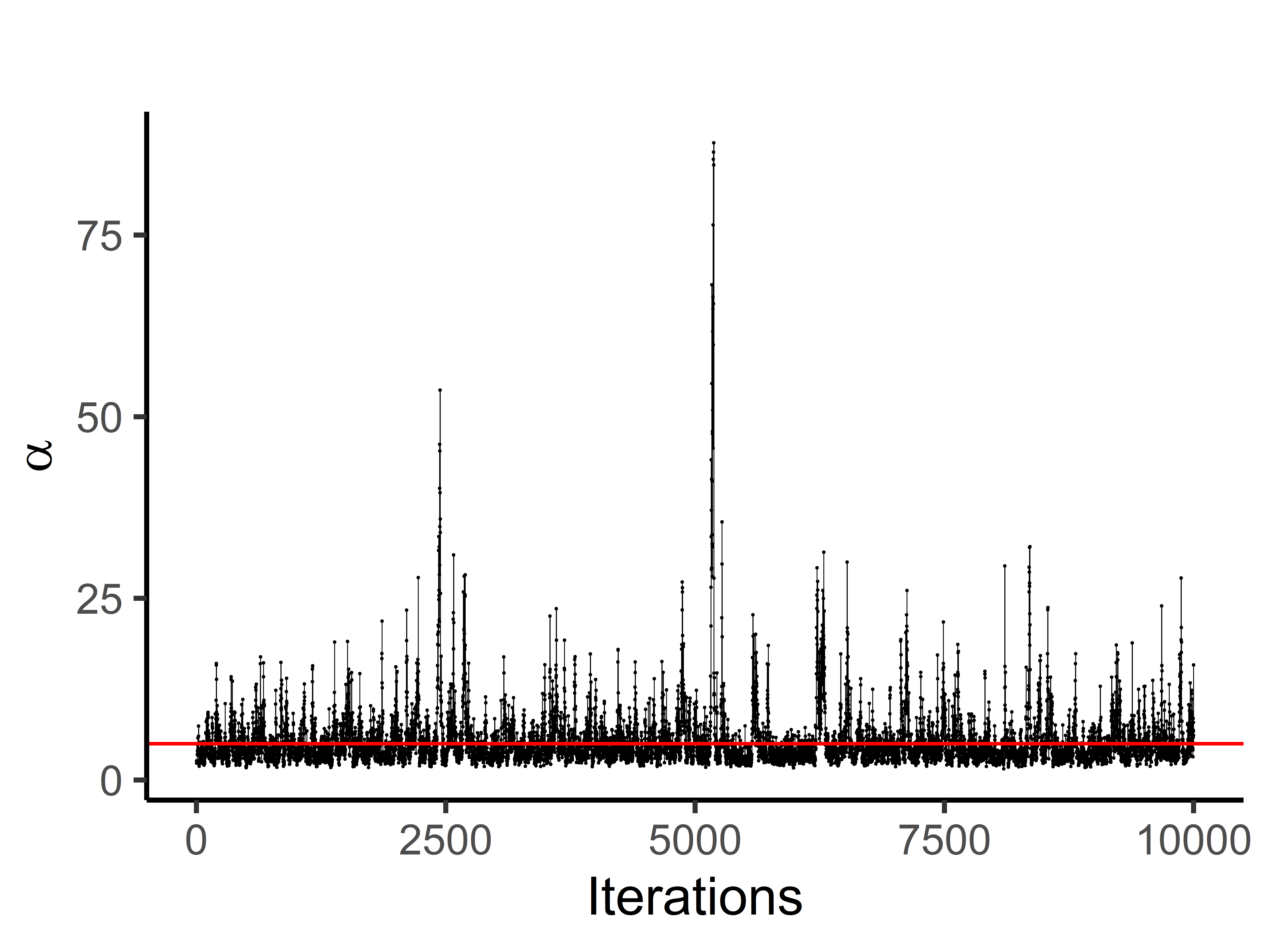} &
   \includegraphics[width=.45\textwidth]{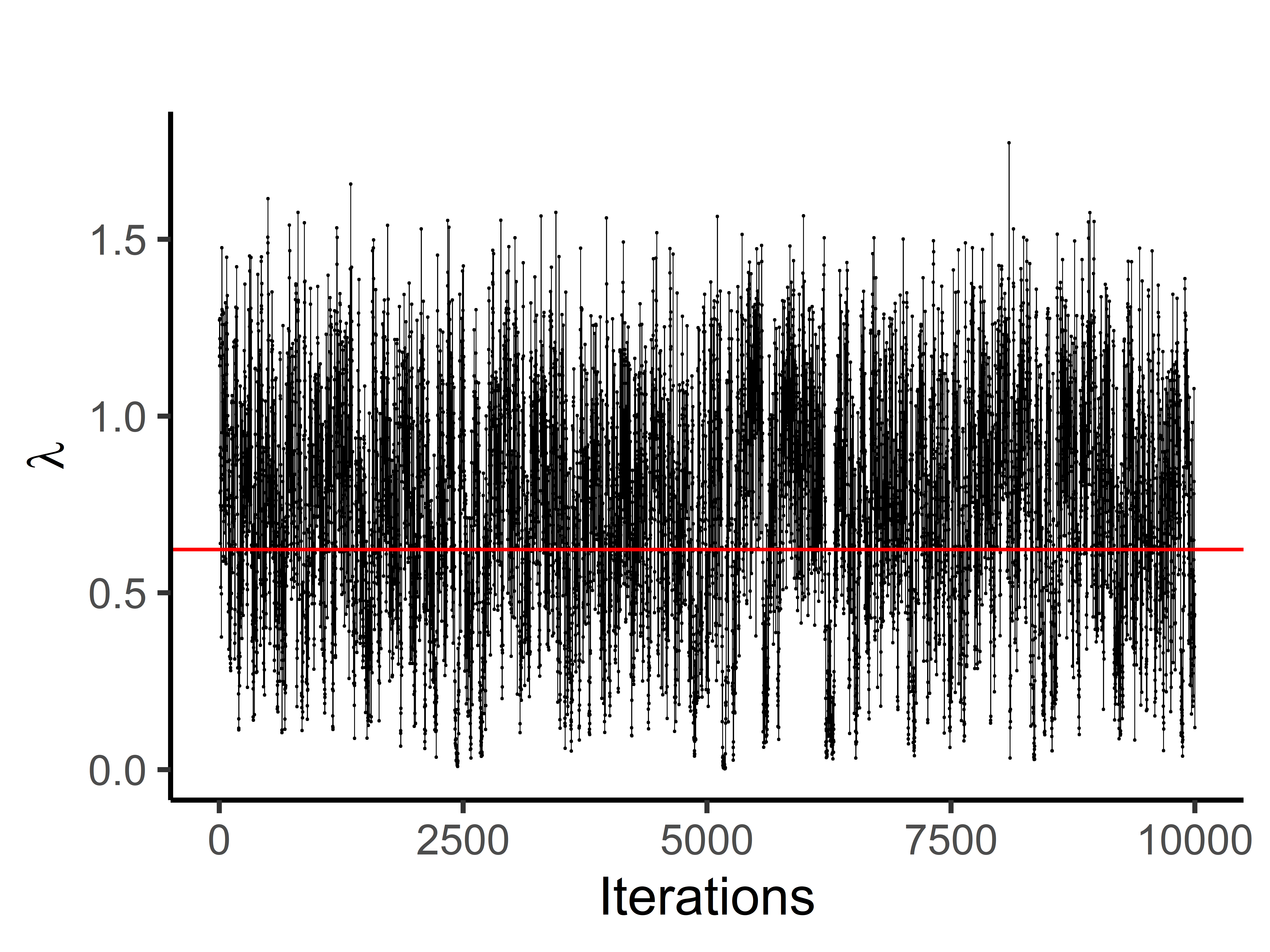}\\
\multicolumn{2}{c}{\includegraphics[width=.45\textwidth]{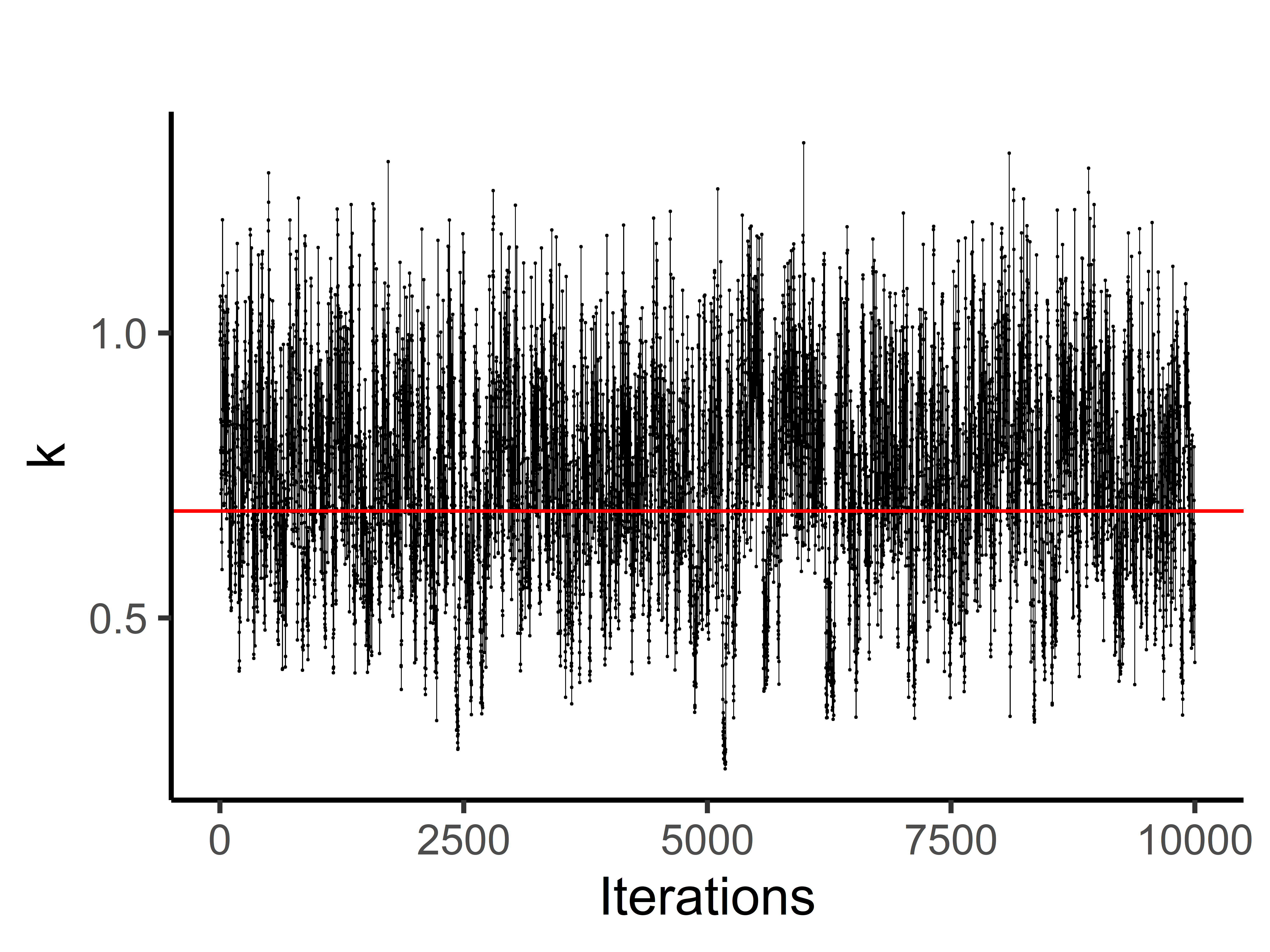}}
  \end{tabular}
  \caption{Parameter estimates progression with respect to iterations in the SEM algorithm }
\label{Fig2}
\end{figure}

{\hspace{7mm}} {\bf {\it Burn-in period}}: For the real data set, 10000 iterations of the stochastic EM algorithm are carried out. For each iteration, the SEM estimate for each parameter is plotted against iteration index (Figure \ref{Fig2}). It is observed that all parameters show similar random behavior around the horizontal line with no discernible pattern, except for $\alpha$.  The plot for $\alpha$ though doesn't show upward, downward or any other obvious pattern, yet the variability around the middle horizontal line is large and doesn't show any obvious diminishing trend. This explains the large standard error that we have obtained corresponding to $\alpha$. The middle horizontal lines correspond to the parameter values which return the maximized log-likelihood value after a burn-in period of 5000 iterations. The random oscillation with almost constant variance around the horizontal line indicates convergence of the SEM estimates to a stationary distribution. However, large variability in the estimates of $\alpha$ is a concern and just taking the average over the iterations after the burn-in period results in under-estimated variance. So, it is reasonable to consider the parameter estimates of the SEM algorithm as the one which return the maximized log-likelihood value after the burn-in period (see \citealp{nielsen2000stochastic}). \\

{\hspace{7mm}} {\bf {\it Model discrimination}}: The cutaneous melanoma data set is further analyzed by fitting all nested sub-models of the EW lifetime distribution as mentioned in Section \ref{sect5}.  The parameter estimates and corresponding standard errors are presented in Table \ref{labint3}. To verify the appropriateness of fitting EW lifetime distribution to the melanoma data under mixture cure rate set-up, maximized log-likelihood ($\hat l$) values are calculated for all sub-models and formal hypotheses tests are carried out to test whether the sub-models deviate significantly from the model with EW lifetime distribution. By using the Wilk's theorem, i.e., $$ -2( \hat l_{EW}- \hat l_{sub}) \overset{a.s.}{\sim}  \chi^2_{\zeta},$$ where $\hat l_{EW}$ and $\hat l_{sub}$ are the respective maximized log-likelihood values under the EW model (alternative model) and sub-model (null model), and $\zeta$ is the difference in the number of parameters estimated, respective p-values for all sub-models are obtained (Table \ref{labint3}). The p-values indicate that all nested models, except the one fitted with  generalized exponential distribution, are significantly different from the EW model, and hence are rejected. Further, Akaike information criterion $$AIC=-2\hat l_{fit}+2q$$ values for each fitted model are also presented in the same table where $\hat l_{fit}$ is the maximized log-likelihood value under the fitted model and $q$ denotes the number of parameters estimated. AIC values suggest that the generalized exponential (AIC=1036.642) model provides the best fit. Hence, for the considered cutaneous melanoma data, the EW lifetime distribution reduces to the generalized exponential distribution. Note the closeness of the generalized exponential model to the EW model based on the AIC values.

\begin{table}[hptb!]
    \centering
   \caption{A comparison of model fitting and inferential results among nested sub-models of the EW lifetime distribution for the cutaneous melanoma data}   
   \begin{tabular}{c|cccccc}
\hline
&\multicolumn{6}{ c }{Fitted Models }\\   
\cline{2-7}
Measure  & EW& Exp& Ral&W&GE&Burr\\ 	
\hline
$\hat \beta_0$(SE($\hat \beta_0$))&-1.117(0.283)&-0.936(0.341)&	-1.200(0.260)&	-1.154(0.265)&	-1.151(0.269)&	-1.168(0.263)\\
$\hat \beta_1$(SE($\hat \beta_1$)) & 0.495(0.113)&	0.607(0.159)&	0.460(0.100)&	0.463(0.103)&	0.478(0.105)&	0.460(0.102)\\
$\hat \alpha$(SE($\hat \alpha$)) &	5.009(4.704)&	- &	- &	- &	2.441(0.316)&	0.764(0.074)\\
$\hat \lambda$(SE($\hat \lambda$)) &	0.623(0.657)&	3.281(0.549)&	2.218(0.097)&	2.217(0.135)&	1.239(0.141)&	2.512(0.181)\\
$\hat k$(SE($\hat k$)) & 0.688(0.308)&	- &	- &	1.616(0.105)&- 	&-	\\
$\hat l$ & -513.839&	-535.037&	-523.897&	-517.593&	-514.321&	-519.932\\
AIC &		1037.678&	1076.074&	1053.794&	1043.186&	1036.642&	1047.864\\
p-value &	-	& $6.221\times 10^{-10}$&	$4.284\times 10^{-5}$&	0.023&	0.617 &	0.002\\
\hline
    \end{tabular}
    \label{labint3}\\
\raggedright \footnotesize{$\hat l$: Maximized log-likelihood value, SE: standard error, EW: Exponentiated Weibull, Exp: Exponential, Ral: Rayleigh, W: Weibull, GE: Generalized Exponential,  AIC: Akaike Information Criterion}\\
\end{table}

\section{Concluding remarks} \label{sect7}

{\hspace{7mm}} The main contribution of this manuscript is the development of the SEM algorithm in the context of Bernoulli cure rate model when the lifetimes of the susceptible individuals are modeled by the EW family of distributions. Different approaches of computing the estimates under the SEM framework have been discussed. An extensive Monte Carlo simulation study demonstrates the accuracy of the SEM algorithm in estimating the unknown model parameters. When compared with the well-known EM algorithm, we have shown that the proposed SEM algorithm is more robust to the choice of initial values than the EM algorithm. This can be seen as an advantage of the SEM algorithm over the EM algorithm. As far as the robustness with respect to outliers is concerned, we have seen that both SEM and EM algorithms perform similar. Hence, in this case, one cannot be preferred over the other. A detailed model discrimination study using the likelihood ratio test clearly shows that different sub distributions of the EW distribution can be easily discriminated. Hence, blindly assuming a distribution for the lifetime is not recommended. Through the real cutaneous melanoma data, we have illustrated the flexibility of the proposed EW distribution. In this regard, we have seen that the assumption of the EW distribution allows formal tests of hypotheses to be performed to select the generalized exponential distribution as the best fitted distribution. In particular, we have seen that all other special cases of the EW distribution get rejected. As potential future works, we can develop the SEM algorithm under a semi-parametric framework, where the lifetimes of susceptible individuals are modeled using the proportional hazards structure and the baseline hazard is approximated using either a piecewise exponential function or a piecewise linear function. This will relax the assumption of homogeneity of lifetimes and the likelihood inference will not depend on any distribution assumptions. Another direction will be to consider more complicated cure rate models such as the ones that look at the elimination of risk factors after an initial treatment and study the performance of the SEM algorithm. We can also think of extending the current framework, as studied in this manuscript, to accommodate interval censored data, as opposed to the commonly used right censored data. We are currently investigating some of these open problems and we hope to report our findings in future manuscripts.

\bibliographystyle{apacite}
\bibliography{ewsem}

\end{document}